\documentclass[12pt,letterpaper,onecolumn]{article}
\usepackage[utf8]{inputenc}
\usepackage[T2A, T1]{fontenc}
\usepackage[russian, english]{babel}
\usepackage{natbib}
\usepackage{setspace}
\usepackage{fancyvrb}
\usepackage{tikz}
\usepackage{etex}
\usepackage[utf8]{inputenc}
\usepackage{natbib}
\usepackage{setspace}
\usepackage{graphicx}
\usepackage{natbib}
\usepackage{amsfonts}
\usepackage[reqno]{amsmath}
\usepackage{amssymb,enumerate}
\usepackage{dcolumn}
\usepackage{tabularx}
\usepackage[left=2.5cm,right=2.5cm,top=3.5cm,bottom=3.5cm]{geometry}
\usepackage{url}
\usepackage{longtable}
\usepackage{booktabs}
\usepackage{rotating}
\usepackage{soul,color}

\makeatletter
\renewcommand{\section}{\@startsection{section}{1}{0mm}{-\baselineskip}{0.25\baselineskip}{\centering\normalfont\normalsize\scshape}}
\renewcommand{\subsection}{\@startsection{subsection}{2}{0mm}{-\baselineskip}{0.25\baselineskip}{\raggedright\normalfont\normalsize\itshape}}
\renewcommand{\subsubsection}{\@startsection{subsubsection}{3}{0mm}{-\baselineskip}{0.25\baselineskip}{\raggedright\normalfont\small\scshape}}
\def\@begintheorem#1#2{\trivlist \item[\hskip \labelsep{\bf #1\ #2:}]\it}
\makeatother
\title{Latent Estimation of GDP, GDP per capita, and Population from Historic and Contemporary Sources}
\author{Christopher J. Fariss  \\ Department of Political Science \\ University of Michigan \\ \texttt{cjf0006@gmail.com} \thanks{Contact Author. We thank Pablo Barber\'a, Michael Beckley, Kristian Gleditsch, James Lo, and Mark Major for helpful comments on this project. All data and code necessary to replicate the model and predicted data presented in the article will be publicly available upon publication at dataverse repositories maintained by the authors. This research was supported in part by The McCourtney Institute for Democracy Innovation Grant, and the College of Liberal Arts, both at Pennsylvania State University and the Security and Political Economy (SPEC) Lab at the University of Southern California.}
\and
Charles D. Crabtree \\ Department of Political Science \\ University of Michigan \\ \texttt{ccrabtr@umich.edu}
\and
Therese Anders \\ School of International Relations \\ University of Southern California \\ \texttt{tanders@usc.edu}
\and
Zachary M. Jones \\ Department of Political Science \\ Penn State University \\ \texttt{zmj@zmjones.com}
\and
Fridolin J. Linder \\ Department of Political Science \\ Penn State University \\ \texttt{fridolin.linder@gmail.com}
\and
Jonathan N. Markowitz \\ School of International Relations \\ University of Southern California \\ \texttt{jnmarkowitz@gmail.com}}
\date{}

\begin{document}

\maketitle

\newpage
\doublespacing

\singlespace
\begin{abstract}
The concepts of Gross Domestic Product (GDP), GDP per capita, and population are central to the study of political science and economics. However, a growing literature suggests that existing measures of these concepts contain considerable error or are based on overly simplistic modeling choices. We address these problems by creating a dynamic, three-dimensional latent trait model, which uses observed information about GDP, GDP per capita, and population to estimate posterior prediction intervals for each of these important concepts. By combining historical and contemporary sources of information, we are able to extend the temporal and spatial coverage of existing datasets for country-year units back to 1500 A.D through 2015 A.D. and, because the model makes use of multiple indicators of the underlying concepts, we are able to estimate the relative precision of the different country-year estimates. Overall, our latent variable model offers a principled method for incorporating information from different historic and contemporary data sources. It can be expanded or refined as researchers discover new or alternative sources of information about these concepts. 
\end{abstract}

\doublespace

\clearpage

\section*{Introduction}
The concepts of Gross Domestic Product (GDP), GDP per capita (GDPpc), population, or some combination of the three rest at the heart of many political and economic theories of interstate and intrastate behaviors.  %For example, John Maynard Keynes employed GDP as a tool to not only measure the economic output of a state but also to predict geopolitical outcomes such as which side would win World War II. 
Since GDP, population, and GDP per capita are strongly related to many important factors, social scientists often include measures of these concepts in their empirical analyses. In some cases, researchers are substantively interested in estimating the effect of these variables on an outcome of interest (e.g., \citealt{gilpin1981,kennedy1987,PrzeworskiAlvarezCheibubLimongi:2000,charron2010does}). Others make the explanation of the growth of population, GDP, or GDP per capita the focal point of their research (e.g., \citealt{north1990,mankiwetal1992,galor2005,acemoglurobinson2012}). In many other cases, researchers include these measures to control for cross-country differences in population or the size of the economy.\protect\footnote{This occurs despite the concern that including these variables in statistical analyses could introduce post-treatment bias, as they might be measured after the outcome of interest has occurred \citep{przeworski2007science}. \citet{king2006publication} points out, ``this is an important point that is often missed'' (124). The consequence of this bias is that it can prevent causal interpretations of estimated coefficients \citep{king2007can}.}  Yet, despite the vital role that these variables play in empirical social science, there are theoretical reasons to believe that they are measured imprecisely. Indeed, a growing literature suggests that existing measures of these important quantities contain considerable error \citep{jerven2010relativity,jerven2010random,jerven2010african,jerven2013poor,Jerven:2014,wb2014,wallace2016juking}. These errors could substantially affect the inferences we make about these variables on other social phenomena either directly or indirectly \citep{stefanski1985covariate,Wooldridge:2010}.

In this paper, we address this construct validity issue by creating and making available new estimates of GDP\protect\footnote{For observed data on GDP see \citet{wb2016, broadberryklein2012, maddison2010, gleditsch2002, bairoch1976}.}, GDP per capita\protect\footnote{For observed data on GDP per capita see \citet{wb2016,broadberry2015,maddison2013,broadberryklein2012,gleditsch2002,bairoch1976}.}, and population.\protect\footnote{For observed data on population see \citet{wb2016, broadberryklein2012, maddison2010, gleditsch2002, singeretal1972}.} The estimates are obtained by combining multiple sources of information for each of these concepts in a dynamic, three-dimensional latent trait model. These measures improve on alternatives in several important ways. First, the dynamic latent variable model extends the temporal and spatial coverage of existing measures by combining information from contemporary and historical sources. The resulting estimates extend from 1500 A.D. to 2015 A.D., the most recent year the observed data are available.\protect\footnote{A country enters the dataset in 1500 A.D. or the first year in which at least one of the datasets records a value for at least on of the included variables (e.g., England 1500-2015 A.D.; Argentina 1800-2015 A.D.; Ghana 1820-2015 A.D.).} Second, because the estimates make use of multiple indicators of the underlying concepts, we are able to estimate the relative precision of the different country-year values for each of the observed measures included in the model. The model provides a complete set of historical estimates for these three latent variables in the form of prediction intervals for thousands of previously missing country-year units in addition to prediction intervals for the original observed country-year variables themselves.\protect\footnote{Missing values are inferred from the dynamic structure of the model based on the information contained in the variables associated with a given country year and the estimates from prior and future values.} Information from these country-year prediction intervals can be incorporated into empirical models that estimate the relationship between these variables and other variables of historic or contemporary interest. Finally, because the models are used to generate predictive intervals for the original source data, the new estimates are available for scholars in the unit-of-measurement of the original observed variables. The resulting dataset covers 35,299 country-year units, for which 16,145 of these country-year units were previously missing. Overall, because the latent variable model offers a principled method for incorporating information from different historic and contemporary data sources and estimating unit specific uncertainty, it can be expanded or refined as other researchers discover new or alternative data sources of these concepts or other information about the recording and reporting processes themselves. We demonstrate this feature of the model using historic data for Ghana.

Before discussing the latent variable model and its estimates, we review the known sources of error in existing measures. Next we describe the GDP, GDP per capita, and population variables. We then introduce the dynamic measurement model and describe how it incorporates information from these three sets of variables. After that, we assess the construct validity of the posterior predictive intervals by comparing them to the observed variables. We close with a discussion of several possible extensions of the model.

\section*{Sources of GDP, GDP per capita, and Population Measurement Error}
Existing measures of GDP, GDP per capita, and population attempt to quantify three related constructs. Gross Domestic Product (and the related Gross National Product) are intended to measure the total amount of the economic output created in a country in a given year \citep{wb2016}. Population measures are intend to reflect the total number of people within the geographic unit in a given year \citep{wb2016}. GDP per capita (and the related GNP per capita) is a measure of the total amount of the economic output created in a country per person in a given year  \citep{wb2016}. The empirical constructs that are intended to map on to these theoretical concepts suffer from a fundamental measurement problem \textemdash\, they are incompletely observable and must be inferred from economic source material and census information. It is extremely labor intensive and cost prohibitive to count every individual within most administrative units (e.g., municipalities, counties, provinces, states) or to track every unit of wealth created in a country (GDP) or by a country's firms (GNP) each year. This means that these empirical constructs can only be \emph{estimated} each year and that some degree of error will always be present in the resulting variable values available in a country's available statistical material. While there must be some error in how they are measured, a large and growing literature suggests that current measures contain considerable error and that it is possibly even systematically related to the estimation process itself (a point we return to in the conclusion) \citep[e.g.,][]{jerven2010relativity,jerven2010random,jerven2010african,jerven2013poor,Jerven:2014,wallace2016juking}.

%\subsection*{Shared Sources of Error}
As \citet{jerven2013poor} shows, there are many reasons for this measurement error. One source of error is that states have strategic incentives to release biased statistics \citep{wallace2016juking}. On one hand, states might have reason to over-report official statistics. This is because high population and GDP numbers might increase domestic support, impress the governments of other states, or gain the attention of foreign investors. The Chinese government, for example, appears to exaggerate its official estimates to forestall opposition criticism and thereby improve regime support \citep{wallace2016juking}. A similar logic allegedly motivates the Belarusian government to release what are known to be very conservative emigration estimates \citep{Yeliseyeu:2014}. The widely circulated rumor is that state leaders do not want the public to know that many Belarusians find other countries more attractive homes. States might also under-report official statistics. This is because low population and GDP numbers might attract foreign aid \citep{neumayer2003,burnsidedollar2000} or other forms of international assistance \citep{buetheetal2012}. These strategic incentives might lead to systematic reporting errors in the source material used to construct the data \citep{Fariss:2014, Scott:1998}.

Another issue is that states do not, or cannot, collect accurate data. Even countries with a well developed census bureaucracy face political issues in accurately counting all groups with the same level of accuracy \citep{Prewitt:2010}.  The situation is worse in some parts of the developing world, where population estimates are often outdated simply because many countries do not or cannot conduct a regular census \citep[e.g.,][]{DriscollLidow:2014}.  The absence of a regular census might be because states are not interested in gathering these data or because they do not have the capacity to do so. Regardless of the cause, the result is the same \textemdash\, outdated data for many country-year observations.

Similarly, GDP estimates are often outdated because countries do not update their `base year' regularly \citep{DQI:2016}. In certain years, statisticians collect additional data on the structure of a country's economy \citep{jerven2013rebase}. These years, called `base years,' are then used to weigh the relative contribution of those sectors to the national income \citep{jerven2013poor}. The contribution of sectors can change considerably over time, as new industries emerge and old ones disappear \citep[141-144]{jerven2013rebase}. As a result, the United Nations and International Monetary Fund both recommend that states `rebase' their estimates every five years to capture changes in the influence of sectors within countries \citep{jerven2013rebase,DQI:2016}. Many countries, however, update this quantity less frequently than suggested. For example, a recent survey suggests that only $7$ of $48$ countries in sub-Saharan Africa follow this recommendation  \citep[146]{jerven2013rebase}. When these countries do eventually update this quantity, it can lead to substantial changes in GDP estimates. For example, Nigeria garnered attention in 2013 when an update to its base year resulted in an 89 percent increase of its GDP estimate for that year \citep{economist2014nigeria}. As this case illustrates, using old base years can cause considerable error in reported GDP estimates.

%*{Sources of Error for GDP}
In addition to these shared sources of error, there are many other factors that specifically influence the accuracy of GDP estimates. One such factor relates to the United Nations System of National Accounts (SNA or UNSNA) \citep{UN:2008}. The SNA is a set of international recommendations about measuring economic activity, designed to make cross-national comparisons easier \citep{deatonheston2010,DQI:2016}. The recommendations are updated semi-regularly (e.g., \citep{sna1953,UN:2008}). However, not all states use these standards and many that do might not use the most up to date version \citep{DQI:2016}. This leads to considerable heterogeneity across countries in the degree to which reported GDP statistics capture underlying economic activity as reflected by a standardized reporting procedure.

Another factor that can lead to misreporting is the size of a country's informal, or shadow, economy. This sector of the economy comprises a mix of legal activities, such as self-employment and bartering, and illegal activities, such as drug-dealing and gambling \citep[5]{Causes2000}. In developing countries, the informal market can represent a substantial share of the economy \textemdash\, perhaps more than 40 percent \citep{ILO:2013}. Even in OECD countries, it can represent about 15 percent of all activity \citep{ILO:2013}. While states usually try to incorporate informal work in their national statistics, this can be difficult since it requires accurately calculating the value of this work \citep{Causes2000}. %As a result, a seventh to a half of all economic activity within countries is measured with at least some uncertainty.

A third additional factor that can lead to error in GDP estimates is whether and how states account for public sector outputs \citep{atkinson2005}. There are two issues here. One is that the national accounts system in most states only tracks government inputs, such as salaries and fixed costs, but not government outputs, such as education and welfare services \citep[31]{stiglitzetal2009}. By not accounting for government outputs, states underestimate national economic activity \citep{stiglitzetal2009}. Another issue is that governments often subsidize the cost of public goods, leading to sub-market prices for these goods \citep[17]{stiglitzetal2009}. As a result, states that measure government outputs based on their market value also underestimate national economic activity.

Despite these sources of error, and the potential dangers they pose to empirical work, there have been few attempts to systematically account for them across all available data sources.\protect\footnote{These has been substantial work on understanding the error in government statistics for a subset of countries and years, particularly in countries in Africa \citep[e.g.,][]{jerven2010african,jerven2010random,jerven2010relativity,jerven2013poor,Jerven:2014}.} Most efforts to improve population and GDP measures have focused on extending their spatial or temporal coverage (which we are interested in as well). These projects typically `fill in' gaps from country-year series that were the result of conflict or political instability \citep{gleditsch2002}. Perhaps the most prominent example of this is the historic GDP and population dataset constructed by Angus Maddison \citep{maddison2010, maddison2013}. While these efforts have been extremely useful, as reflected in the thousands of citations they have received, they do not directly address our primary concern, which is the existence of bias and error in reported country-year estimates. While this line of research has focused on providing researchers with more data, we add to this contribution by both extending the coverage of the estimates temporally and spatially and also by providing researchers with a means of understanding the relative precision of the resulting estimates. 

Several organizations, such as World Economics \citep{DQI:2016}, have tried to address the concern we raise about these statistics by releasing datasets that measure their quality. There are two primary problems with these datasets, however. The first is that their temporal range is usually extremely limited. The second is that while these measures provide some information about the relative uncertainty of official GDP statistics, they do not incorporate this information into country-year estimates. Our new model and model-based estimates, which we describe below, provide a method for addressing these limitations. Importantly, our model should be considered a useful starting point that can be extended as new information becomes available. We discuss several such extensions and provide one demonstration using information from the case of Ghana \citep{Jerven:2014}.

%\clearpage
\section*{GDP, Population, and GDPpc Component Datasets}
Our latent variable model is estimated based on data for Gross Domestic Product (GDP)\protect\footnote{For observed data on GDP see \citet{wb2016, broadberryklein2012, maddison2010, gleditsch2002, bairoch1976}.}, GDP per capita\protect\footnote{For observed data on GDP per capita see \citet{wb2016,broadberry2015,maddison2013,broadberryklein2012,gleditsch2002,bairoch1976}.}, and population.\protect\footnote{For observed data on population see \citet{wb2016, broadberryklein2012, maddison2010, gleditsch2002, singeretal1972}.} Details on the sources, measurement choices, and coverage of the component variables are provided in Table \ref{tab_ALL}. %\ref{tab_gdp} through \ref{tab_pop}.
For each component dataset, we extract relevant indicators, attach unique country identifiers, and reshape the data into a common country-year format. We consulted the codebooks of each dataset to drop observations that are interpolated or extrapolated by the authors of the dataset, or already covered by other datasets (e.g., the data generated by \citet{gleditsch2002} includes some interpolated values and values taken from the Maddison Project). Details on the underlying source materials for each component measure and coding decisions are provided below and are documented in the \texttt{R} code we use to merge the constituent datasets together.

When merging the different variables together we relied on the available country-year units as prepared by the authors of the original datasets. We use the \citet{GleditschWard:1999} revised list of independent states as the base set of units. For years prior to the start year of this data set (1816 A. D.) we again use the date the year the unit enters the dataset or 1500 A.D. As we discussed in each dataset description, different datasets sometimes use different spatial definitions for units. We have matched country-year units across datasets using the best match available. In some cases, units exist in the dataset that are not historically accurate such as a unified Germany prior to 1871. Maddison includes this unit in his historic data series, aggregating information across the various principalities and other administrative districts that existed until Germany had completely unified in 1871. As another example, Maddison also disaggregates information about North and South Korea backwards in time. Additional details about these unit specific issues are available in the original source material. Documentation about how we merged all of the data sources together are available in our code files, which are publicly accessible. Importantly, because many of these units are subsets of larger ones (e.g., North and South Korea), analysts can aggregate the estimates of these two units together if necessary for a specific empirical application.

% GNP vs GDP
%\vspace{5mm}
%\input{gnp_gdp.tex}

%\clearpage
%\subsection*{Source Material for Component Datasets}\label{sec:sources}
%Source material for the variables are described in detail below and summarized in Table \ref{tab_ALL}.

%%% GDP Table
\clearpage
\singlespace
\begin{center} \scriptsize 
\begin{longtable}{>{\raggedright\arraybackslash}p{6.5cm}>{\raggedright\arraybackslash}p{1.75cm}>{\raggedright\arraybackslash}p{1.75cm}>{\raggedright\arraybackslash}p{5.5cm}} \caption{Component Measures for GDP, GDP per capita, and Population Latent Variable Model}\label{tab_ALL}\\
\toprule
 {\bf Variable Descriptions} & {\bf Coverage in Original} & {\bf Coverage in Model} & {\bf Source Material and Citations}  \\ \midrule
GDP data are measured in 1990 international dollars. & 1AD--2008 & 1500--2008 & Historical GDP data collected by Angus Maddison \citep{maddison2010}. \\ \midrule
GDP data are measured as total real GDP at 2005 prices.  &  1950-2011& 1950-2011& Expanded GDP data version 6.0 beta, September 2014 \citep{gleditsch2002}.\\ \midrule
GDP data are measured in constant 2010 USD. & 1960--2015 & 1960--2015 & World Development Indicators \citep{wb2016} \\ \midrule
GDP data limited to European countries and the United States, after accounting for changing country boundaries. GDP is measured in millions of 1990 international dollars (national currencies are converted to international dollars using Angus Maddison's purchasing power parities) &1870--2001 &1870--2001   & \citet{broadberryklein2012}. \\ \midrule
GNP data limited to European countries, after accounting for changing country boundaries. GNP is measured at market prices and expressed in constant 1960 US dollars.& 1830--1973& 1830--1973 &\citet{bairoch1976}.\\ \midrule

GDP per capita data are measured in 1990 international dollars. &1AD-2010 & 1500--2010 & Extension of Angus Maddison's historical GDP and population estimates \citep{maddison2013}.\\ \midrule
GDP per capita data are measured as total real GDP at 2005 prices.  & 1950-2011& 1950-2011& Expanded GDP data version 6.0 beta, September 2014 \citep{gleditsch2002}.\\ \midrule
GDP per capita are measured in constant 2010 USD. &  1960--2015& 1960--2015 & World Development Indicators \citep{wb2016}\\ \midrule
GDP per capita data limited to European countries and the United States, after accounting for changing country boundaries. GDP is measured in millions of 1990 international dollars. &1870--2001 &1870--2001  & \citet{broadberryklein2012}.\\ \midrule
GNP per capita data are limited to European countries, after accounting for changing country boundaries. GNP is measured at market prices and expressed in constant 1960 US dollars.& 1830--1973& 1830--1973 &\citet{bairoch1976}.\\ \midrule
GDP per capita data limited England/Great Britain, Holland/Netherlands, Italy, Spain, Japan, China, and India. GDP is measured in millions of 1990 international dollars. & 725--1850 & 1500--1850 & \citet{broadberry2015}.\\ \midrule

 Total population measured in thousands at mid-year. &1AD--2030 &1500--2010 & Historical population data collected by Angus Maddison \citep{maddison2010}.\\ \midrule
 Total population measured in thousands. & 1950-2011 & 1950-2011 & Expanded GDP data version 6.0 beta, September 2014 \citep{gleditsch2002}.\\ \midrule
 Population data limited to European countries and the United States. &1870--2001 &1870--2001   & \citet{broadberryklein2012}. \\ \midrule
 Total population. & 1960--2015& 1960--2015 & World Development Indicators \citep{wb2016}\\ \midrule
 Total population measured in thousands.& 1816--2001 & 1816--2001 & The Correlates of War Project's National Material Capabilities data version 4.0 \citep{singeretal1972}
\\ \bottomrule
\end{longtable}
\end{center}
\doublespace

\subsubsection*{The Maddison Project \citep{maddison2010, maddison2013}}
\citeauthor{maddison2010}'s original GDP, GDP per capita, and population variables are derived from a large number of country-level sources \citep{maddison2003, maddison2001, maddison1995}. Because the underlying source materials employed by Maddison are expansive and country-specific, we refrain from describing them in detail. The more recent version of these data, \citet{maddison2013}, is based on a collaboration of researchers dedicated to continuing Angus Maddison's data collection efforts by extending and, if warranted, revising his estimates. Due to the collaborative nature of the effort, different research teams use different methods and source material to obtain their estimates. With a few exceptions, data from 1990--2010 were revised using figures from the Total Economy Database of the Conference Board \citep{boltvanzanden2014}. Other estimates are based on historical national statistics from country-specific sources \citep{boltvanzanden2014}. We subset the data from the Maddison Project to include only country-year observations starting in 1500. The original \citet{maddison2010} data includes both GDP and population values. The updated version only included GDP per capita estimates. We include both data versions in our model since, as we describe below, it is capable of linking all of these observed indicators together in united model that leverages the information from each type of variable. Unlike some of the other datasets we describe below, these datasets do not contain origin codes that classify the source material used to inform the country-year values. 

\subsubsection*{Expanded GDP data version 6.0 beta \citep{gleditsch2002}}
\citet{gleditsch2002}'s (beta) version 6.0 of the Expanded GDP data is based primarily on the Penn World Tables (PWT) 8.0, and supplemented with data from the PWT 5.6, the Maddison Project Database, and the World Bank Global development indicators. In addition, \citet{gleditsch2002} constructed his data using imputations for the lead and tail values, as well as interpolation for estimates within the series. We use only the values that stem from the PWT figures in the latent variable model (origin codes 0, -1, and 3) and exclude data from the Maddison Project, as well as interpolated or imputed figures (origin codes -2, 1, and 2). In the Validity section below, we consider the model fit for the latent variables estimates that do include these variables compared to the latent variable model estimates that exclude them and demonstrate the model fit is improved by estimating these missing values using our model-based approach instead of using interpolation or extrapolation.

\subsubsection*{World Development Indicators \citep{wb2016}}
We include GDP, GDP per capita, and population from the \citet{wb2016}. Where possible, we use the metadata for each indicator provided by the World Bank's DataBank to determine the underlying source material of the GDP, GDP per capita, and population values. As with the \citet{gleditsch2002} data, we drop values that are interpolated or extrapolated and allow our model to generate new estimates for these units. We describe each of these variables in turn.

We include the \citet{wb2016}'s GDP indicator measured in constant 2010 US dollars in our latent variable model. The figures are compiled from the World Bank and OECD national accounts data. The documentation in the metadata file indicates that the series is based on an underlying interpolation of component data upon aggregating it to a ``gap-filled total.'' Unfortunately, we do not have information on the details of this aggregation process. We therefore use the full series of GDP as provided by the \citet{wb2016}'s online data portal DataBank. In future versions of our model, we plan to identify these cases when possible and adjust our model accordingly. 

The per capita GDP series is based on the \citet{wb2016}'s GDP in constant 2010 US dollars and the total population figures (for the underlying source material see below). According to the metadata, the data is aggregated using weighted averages. We exclude observations from our model that the metadata indicates as being preliminary, extrapolated, or interpolated. Information on which country-years were excluded based on the metadata is provided in the replication material that accompanies this paper.

The population figures from \citet{wb2016} are based on national population censuses. The census data that informs this measure stem from a variety of sources, including the United Nations World Population Prospects (for the majority of developing countries), Eurostat (for European countries), and national statistical agencies. The data are interpolated for all years between census years. Since we do not have information on the years that a census was conducted for each country, we retain the interpolated data for the use in the latent variable model. We do, however, exclude population figures that are explicitly indicated as being extrapolated, interpolated, or preliminary in the metadata. Information on which country-years-units were excluded is provided in  the replication material that accompanies this paper. In future versions of our model, we plan to identify the other interpolated cases when possible and again adjust our model accordingly.

\subsubsection*{\citet{broadberryklein2012}}
The GDP, GDP per capita, and population variables in \citet{broadberryklein2012} are limited to European nations, including Russia and Turkey, as well as the United States. A detailed list of underlying source material is available in the paper's appendix \citep[pp. 105]{broadberryklein2012}. For GDP, these sources include the data from \citet{maddison2010}, official national account statistics, and the work of country-expert historians. Data on population are drawn mainly from \citet{mitchell2003} and \citet{maddison2010}, and supplemented with country-specific data from official national statistics and historians. We exclude those country-year observations that are taken from \citet{maddison2010} in our model.

\subsubsection*{\citet{bairoch1976}}
The underlying source material for the data by \citeauthor{bairoch1976} is detailed in the paper's methodological appendix. For GNP, these sources include the work of historians and official national statistics for earlier country-years, as well as OECD figures for years starting in 1950 \citep[329 et seq.]{bairoch1976}. For the 19th century and the year 1900, three-year annual averages are available for every decade starting from 1830 and expressed in 1960 U.S. dollars \citep[286]{bairoch1976}. For the 20th century, data are available for select years between 1913 and 1973 and expressed in 1960 U.S. dollars as well \citep[297]{bairoch1976}. For population, \citeauthor{bairoch1976} relies on United Nations Demographic yearbooks, data from the League of Nations, and national statistical agencies to assemble his data (321). We incorporate all of \citeauthor{bairoch1976}'s estimates in our model, including the ones flagged as having a larger-than-average margin of error (the figures presented in parentheses). The data from \citet{bairoch1976} cover the total and per capita gross \textit{national} product (GNP), not gross \textit{domestic} product (GDP). \citeauthor{bairoch1976}'s definition is based on the United Nations' 1953 System of National accounts \citep{sna1953}. %Since the GNP and GDP are highly correlated (see Table \ref{gnp_gdp}), we estimate a joint latent variable model between \citeauthor{bairoch1976}'s GNP estimates and the GDP data of other sources. 
With the exception of the data from \citet{bairoch1976}, the data on economic size are measured as the gross domestic product (GDP). \citet{bairoch1976} uses gross national product (GNP) instead. While the GNP excludes value added by foreign firms, this measure is highly correlated with GDP, as demonstrated in Table \ref{gnp_gdp} in the Appendix. The correlation between GNP and GDP is quite high, with correlation coefficients between 0.865 and 0.995 for country-year units within the period 1830--1973. The strength of the positive relationship varies over time but rarely falls below 0.9. We anticipate that in future years, the correlation between the two measures should drop as globalization increases and the internationalization of production and investment increases the relevance of the conceptual difference between GNP and GDP. Additional estimates of GNP and GDP from more recent years would help researchers determine how this empirical relationship evolves over time. The evaluation of this distinction is one possible avenue that our new latent variable model opens up for exploration, which we discuss below.

\subsubsection*{\citet{broadberry2015}}
The GDP per capita estimates in \citet{broadberry2015} are based on historical national accounting data that is constructed from documents such as ``government accounts, customs accounts, poll tax returns, parish registers, city records, trading company records, hospital and educational establishment records, manorial accounts, probate inventories, farm accounts, tithe files and other records of religious institutions.'' \citep[5]{broadberry2015}. \citeauthor{broadberry2015} lists the data sources for each country in the main text.\protect\footnote{Pages 6 and 7 contain the underlying source material for Britain, the Netherlands, Italy, and Spain; page 8 contains the data for China, Japan, and India.} As with the Maddison data, we exclude cases for years prior to 1500 from our model.

\subsubsection*{COW National Military Capabilities data v4.0 \citep{singeretal1972}}
The Correlates of War Project provides a variety of country-level estimates including population beginning in the year 1816. For country-years starting in 1919, the population estimates by \citet{singeretal1972} are based primarily on the estimates of the United Nations Statistical Office. The population estimates for years prior to 1919 are based on national government censuses. For these earlier years in the series, the authors of the population dataset selected country-specific data that presents the greatest continuity with the data from the United Nations.\protect\footnote{For details, please refer to the codebook for version 4.0 of the data: Correlates of War Project National Material Capabilities Data Documentation Version 4.0, \url{http://cow.dss.ucdavis.edu/data-sets/national-material-capabilities/nmc-codebook/at_download/file}, accessed 1 December 2016.} The authors of the data use a variety of methods to bridge gaps in the data, including interpolation, regression, and extrapolation. Quality codes for the estimates of the total population figure are specified \textemdash\, indicating whether a data point stems from an identified source, is missing, derived through interpolation, regression, or extrapolation. We retain only those data points that stem from an identified source (quality code A).

%\clearpage
\section*{Model Description, Specification, and Estimation}

The empirical goal for this paper is to specify a dynamic latent variable model that will allow us to leverage the available information across different contemporary and historical sources of data on GDP, population, and GDP per capita. Unlike other latent variable models, we are not directly interested in the estimated latent traits. Instead we use these estimated traits to generate prediction intervals for each of the observed variables in their original unit-of-measurement. The model therefore allows us to leverage the information across all of the observed data and to generated predictions for each of the individual observed variables. In this way, interested users can select their preferred dataset for empirical applications in conjunction with the prediction intervals generated from our model. Importantly, the prediction intervals are available for every unit included in the dataset. As we discuss later in the manuscript, the size of the prediction interval is related to total coverage and agreement of the observed variables.

To specify the dynamic latent variable model, let $i = 1, \dots, N$, index cross-sectional units and $t = 1, \dots T$, index time periods. For each country-year unit, $j = 1, \dots, J$ indexes the observed variables $y_{itj}$. Because the observed variables that enter the model represent three different concepts --- GDP, population, and GDP per capita --- we estimate three latent variable parameters, where $k = 1, 2, 3$, indexes the three categories $gdp,~pop,~gdppc$. This allows us to define the set of $y_{itj}$ that we observe for each of the $k$ dimensions of the latent variable model, where $y_{itj} * 1 \{ y \in \pi_{k} \}$.  This notation allows us to denote the set of observed variables used to estimate each of the three underlying latent variables such that $\pi_{gdp} = \{ y_{it1}, y_{it2}, y_{it3}, y_{it4}, y_{it5} \}$, $\pi_{pop} = \{ y_{it6}, y_{it7}, y_{it8}, y_{it9}, y_{it10} \}$, $\pi_{gdppc} = \{ y_{it11}, y_{it12}, y_{it13}, y_{it14}, y_{it15}, y_{it16} \}$.\protect\footnote{A useful feature of this notation is that the sets of observed variables do not need to be mutually exclusive. Though we do not allow the observed variables to inform the estimation of multiple latent variables in the application presented here, this is a possibility in other applications. See \citet{GelmanHill:2007, ImaiLoOlmsted:2017} for more details. We thank James Lo for this notational suggestion.}

With knowledge of how the observed variables relate to each category $k$, we can denote how the three dimensions of the latent variable relate to them as well. The model assumes that the latent variables take the form: $\theta_{itk} \sim \mathcal{N}(0, 1)$ for all $i$ when $t=1$ (the first year a country enters the dataset).   When $t>1$, the standard normal prior is centered around the latent variable estimate from the previous year such that:  $\theta_{itk} \sim \mathcal{N}(\theta_{it-1,k}, \sigma_{k})$.

The latent variables themselves are estimated with uncertainty. The first year each country enters the model, the variances for these parameters are set to 1. For all years after $t=1$, $\sigma_{gdp}$ and $\sigma_{pop}$ are drawn from a uniform distribution $U(0,1)$. For the latent GDP per capita variable, the latent estimates and associated uncertainty are deterministically determined by the GDP and Population latent variables themselves such that  $\theta_{it,gdppc}$ $\leftarrow$ $\frac{\theta_{it,gdp}}{\theta_{it,pop}}$. This modeling innovation allows information form the three types of observed variables to inform more than just one of the latent variables.

The latent variables are estimated by linking each of these parameters to the sets of observed GDP, population, or GDP per capita variables. Since all of the GDP, population, and GDP per capita variables are continuous, we specify a Gaussian link function with a unique error term for each of the the three types of variables $\tau_{k}$: $\{ \tau_{gdp}$, $\tau_{pop}$, $\tau_{gdppc} \}$. These $\tau_{k}$ parameters are estimates of model level uncertainty, which link each of the latent variables to the sets of observed GDP, population, or GDP per capita variables.  Shape parameters translate the observed variables from their original unit-of-measurement into the latent variable unit-of-measurement. Because we specify a Gaussian link function, these shape parameters are the intercept and slope from the linear model. For the intercept parameters $\alpha_{j}$, we center the standard normal prior around the the mean value of the observed data with a relatively large variance (low precision): $\alpha_{j}$ $\sim$  $\mathcal{N}(\bar{y}_{j}, 4)$. We choose the mean value of the observed variables because the mean of latent traits themselves are centered around 0.\protect\footnote{We set this parameter to the empirical mean of the Maddison GDP and population variables as an identification constraint.} The intercept parameter therefore transforms the latent trait into the unit-of-measurement of the original observed variable. For identification of the model we set $\beta_{j}=1$ because we assume a one-unit change in the latent trait is equivalent to a one-unit change in the original observed variable.\protect\footnote{This assumption can be relaxed to examine the relative strength of the relationship between one measure compared to another. We leave this analysis to future research. Relaxing this assumption would allow for analysts to explore the relative relationship between measures of GDP and GNP as functions of the underlying latent trait. We view this as a useful extension to the model we present here.}
%The slope parameters $\beta_{j}$ are constrained by truncation so that they are each positive $\beta_j$  $\sim$  $\mathcal{N}(0, 4) \mathcal{T}(0,1)$.
All of the prior distributions are summarized in Table \ref{priors}. Recall that we organize the three types of observed variables in three sets such that $y_{itj} * 1 \{ y \in \pi_{k} \}$. Therefore, the likelihood function that links the observed data to the estimated parameters is:

\[
%\mathcal{L}(\beta, \alpha, \tau, \theta | y * 1 \{ y \in \pi_{k} \} ) = \prod_{i=1}^N \prod_{t=1}^T \prod_{j=1}^J \mathcal{N}(\alpha_{j} + \theta_{itk} \beta_{j}, \tau_{k})
\mathcal{L}(\beta, \alpha, \tau, \theta | y_{itj} * 1 \{ y \in \pi_{k} \}  ) = \prod_{i=1}^N \prod_{t=1}^T \prod_{j=1}^J \prod_{k=1}^K \mathcal{N}(\alpha_{j} + \theta_{itk} \beta_{j}, \tau_{k})
\]

\begin{table}[htbp]
\small
\caption{Prior Distribution for Latent Variables and Model Level Parameter Estimates }
\begin{center}
\begin{tabular}{| l | r c l |}
\hline
Parameter & Prior &  & \\
\hline
Country $i$ latent GDP estimate in first year $t$ & $\theta_{it=1, gdp}$ &$\sim$&  $\mathcal{N}(0, 1)$  \\
Country $i$ latent GDP estimate in all other years & $\theta_{it, gdp}$ &$\sim$& $\mathcal{N}(\theta_{t-1,gdp}, \sigma_{gdp})$  \\
Latent GDP uncertainty & $\sigma_{gdp}$ & $\sim$ & $U(0,1)$  \\

& & & \\
Country $i$ latent population estimate in first year $t$ & $\theta_{it=1,pop}$ &$\sim$&  $\mathcal{N}(0, 1)$  \\
Country $i$ latent population estimate in all other years & $\theta_{it,pop}$ &$\sim$& $\mathcal{N}(\theta_{t-1,pop}, \sigma_{pop})$  \\
Latent population uncertainty & $\sigma_{pop}$ & $\sim$ & $U(0,1)$  \\

& & & \\
Country $i$ latent GDP per capita estimate & $\theta_{it,gdppc}$ &$\leftarrow$& $\frac{\theta_{it,gdp}} {\theta_{it,pop}}$  \\

& & & \\
Model $j$ intercept ``difficulty parameter'' & $\alpha_{j}$ & $\sim$ &  $\mathcal{N}(\bar{y}_{itj}, 4)$ \\
%Model $j$ slope ``discrimination parameter'' & $\beta_j$ & $\sim$ & $\mathcal{N}(0, 4) \mathcal{T}(0,1)$   \\
Model $j$ slope ``discrimination parameter'' & $\beta_j$ & $\leftarrow$ & $1$   \\

& & & \\
Model uncertainty for all GDP items& $\tau_{gdp}$ & $\sim$ & $\mathcal{G}(0.001, 0.001)$  \\
Model uncertainty for all population items & $\tau_{pop}$ & $\sim$ & $\mathcal{G}(0.001, 0.001)$  \\
Model uncertainty for all GDP per capita items & $\tau_{gdppc}$ & $\sim$ & $\mathcal{G}(0.001, 0.001)$  \\

 \hline
\end{tabular}
\end{center}
\label{priors}
\end{table}%

The model is estimated with five MCMC chains, run for 100,000 iterations each. The first 50,000 iterations were thrown away as burn-in and the rest were used to generate the posterior prediction intervals for the original observed variables.\protect\footnote{The Gibbs sampler was implemented in Martyn Plummer's JAGS software \citep{Plummer:n.d.}. The JAGS code used is displayed in the Appendix. Conventional diagnostics all suggest convergence.}

\clearpage
\section*{Model Validity Checks}
\subsection*{Convergent Validity}
\begin{figure}[htbp!]
    \centering
    \includegraphics[width = 8.5in]{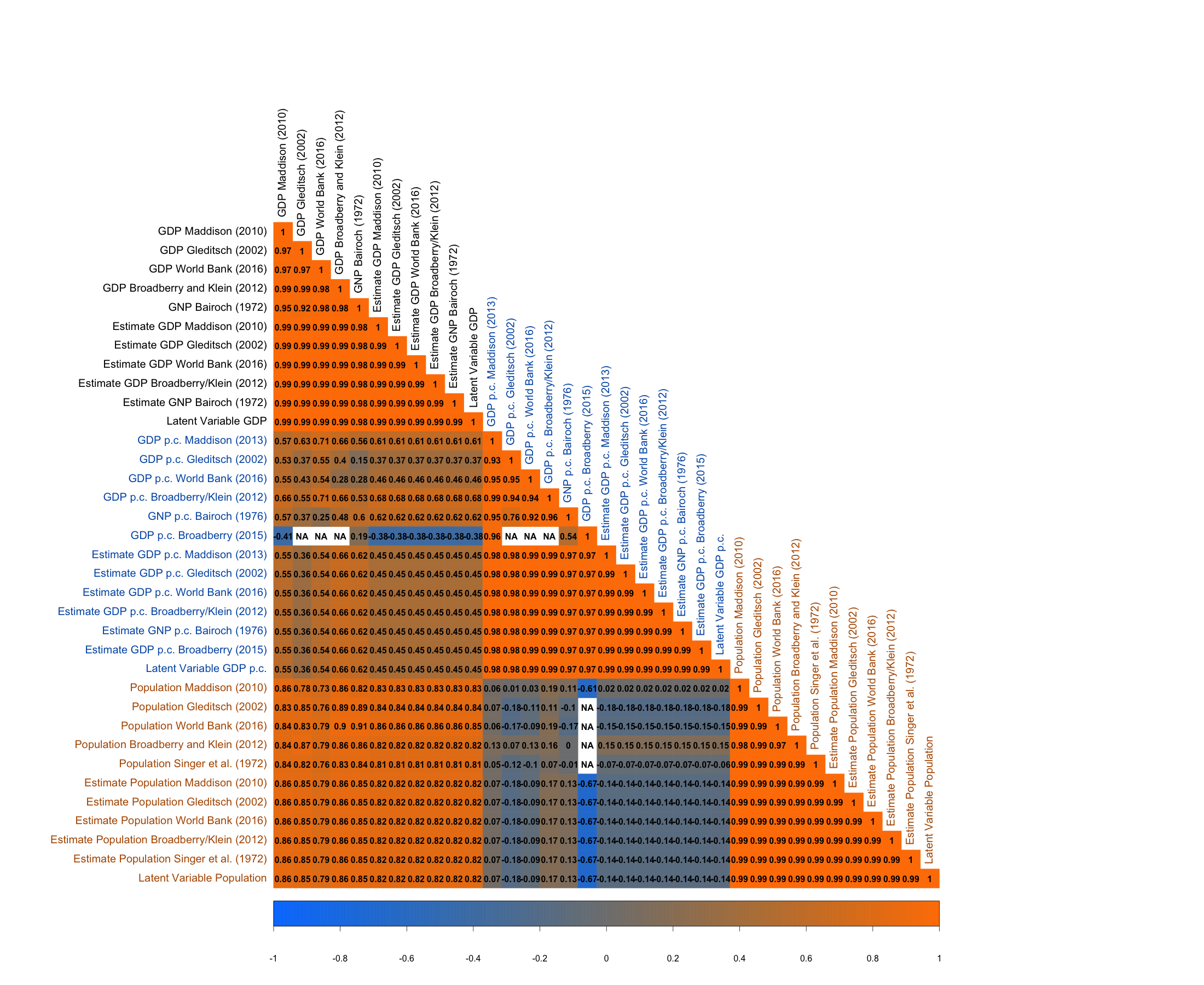}
    \caption{The figure presents the pairwise correlations between all component variables of the latent variable model, their estimates, and the latent variables for GDP, per capita GDP, and population. The shading indicates the value of the correlation coefficient [$-1, 1$]. White cells with a value of NA denote pairings with no overlapping country-year observations. Note that cells that round up to 1 but are not exactly 1 (i.e. all cells that are not in the diagonal), are represented as 0.99.}
    \label{fig:pairwisecorr}
\end{figure}

\begin{figure}[htbp!]
\begin{center}
\includegraphics[width = 6.0in]{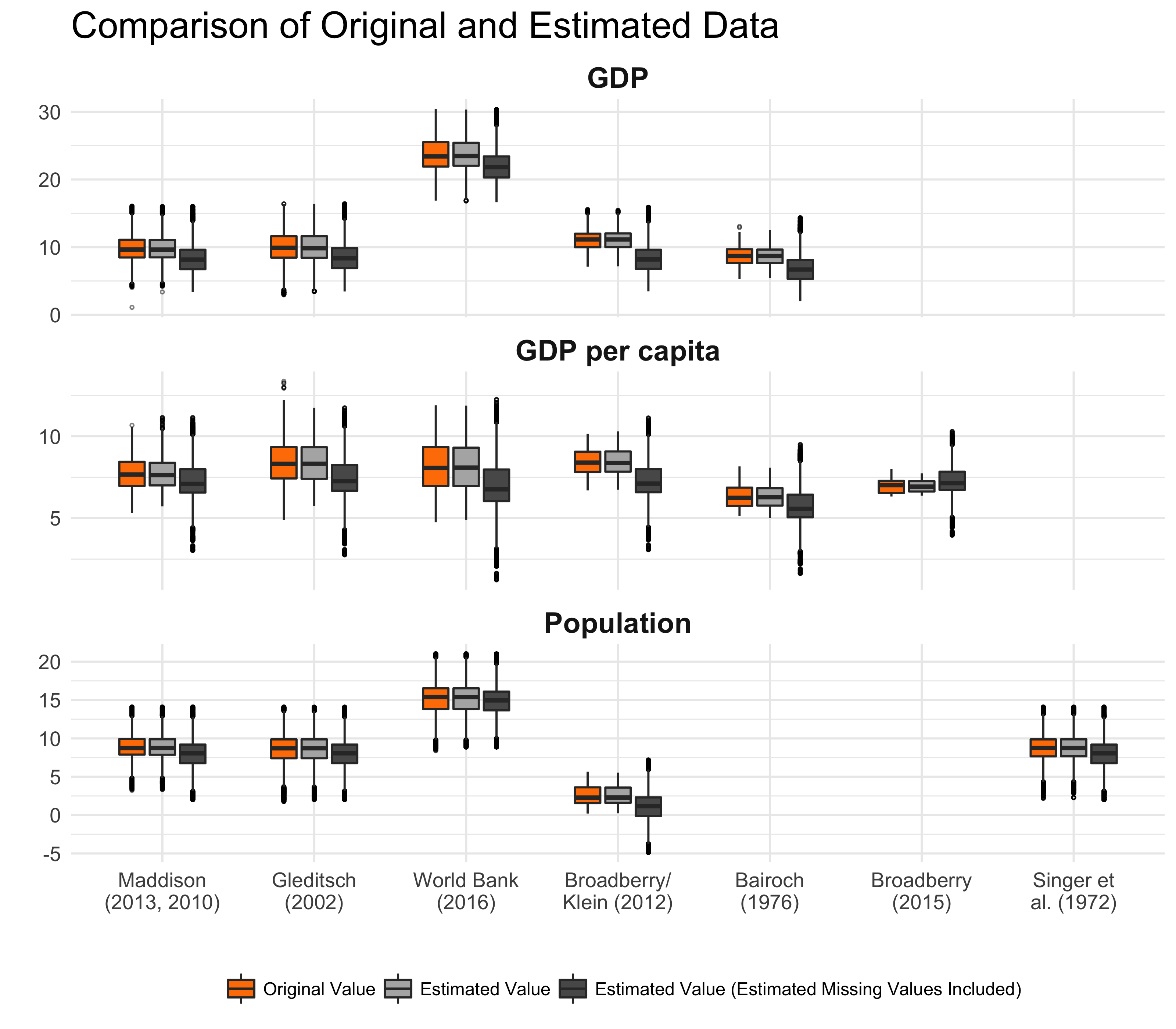}
\caption{The plot illustrates the agreement between the observed country-year-variable values (orange boxes) and the posterior predicted point estimates for which the observed value is missing (light grey boxes). Dark grey boxes show the distribution for the full range of the posterior predicted point estimates (including estimates for which the original value is missing). Across all variables, these estimated values have a lower median value due to a bias in the missingness in the original data. We have more missing observations in earlier years, for which GDP, per capita GDP, and population levels are lower than in later years in the series. Conveniently for users, the posterior predicted values are estimated using the original unit-of-measurement. This means that the visual discrepancy, for example between the \citet{wb2016} and the other population data, is not an empirical discrepancy. It is simply the difference in the unit-of-measurement used for the observed values (i.e., the \citet{wb2016} publishes the absolute GDP and population values while others present these in per thousand units). Table \ref{Proportion_Table} quantifies the similarity between the original data and the full range of possible values along the posterior predicted range, demonstrating that approximately 86 percent of the observed country-year-variable values are within $\pm$ 1 standard deviation of the predicted range of the estimated data.}
\label{plot_original_estimated}
\end{center}
\end{figure}

To assess the validity of the new estimates, we first check for convergent validity. We do this by comparing the posterior prediction intervals and the observed variables we used to estimate the latent traits. Figure \ref{fig:pairwisecorr} displays the relationship between the original observed variables ($y_{itj}$), the mean or expected value of the latent trait estimates ($E(\theta_{itk})$), and the mean or expected value of the posterior prediction intervals ($E(\tilde{y}_{itj})$). The plot demonstrates the high level of agreement between the original observed variables and the new posterior predictions. Figure \ref{plot_original_estimated} illustrates this agreement for each observed variable and its corresponding posterior predictions. Conveniently for users, the posterior predicted values are estimated using the original unit-of-measurement (e.g., the natural log of 1990 International Dollars). This means that the visual discrepancy, for example between the \citet{wb2016} and the other population data, is not an empirical one. Instead, it is simply the difference in the unit-of-measurement used for the observed values (i.e., the \citet{wb2016} publishes the absolute GDP and population values while others present these per thousand individuals).

 \clearpage
% Correlation plot

\clearpage
\subsection*{Precision of Estimates}
As we discussed above, the underlying latent traits themselves are estimated with uncertainty.  Conveniently, the amount of uncertainty for any particular observation is available for each posterior predicted unit. This information should be incorporated into any analysis that utilizes these new estimates \citep{SchnakenbergFariss:2014}. 

\begin{figure}[htbp!]
\begin{center}
\includegraphics[width = 6.0in]{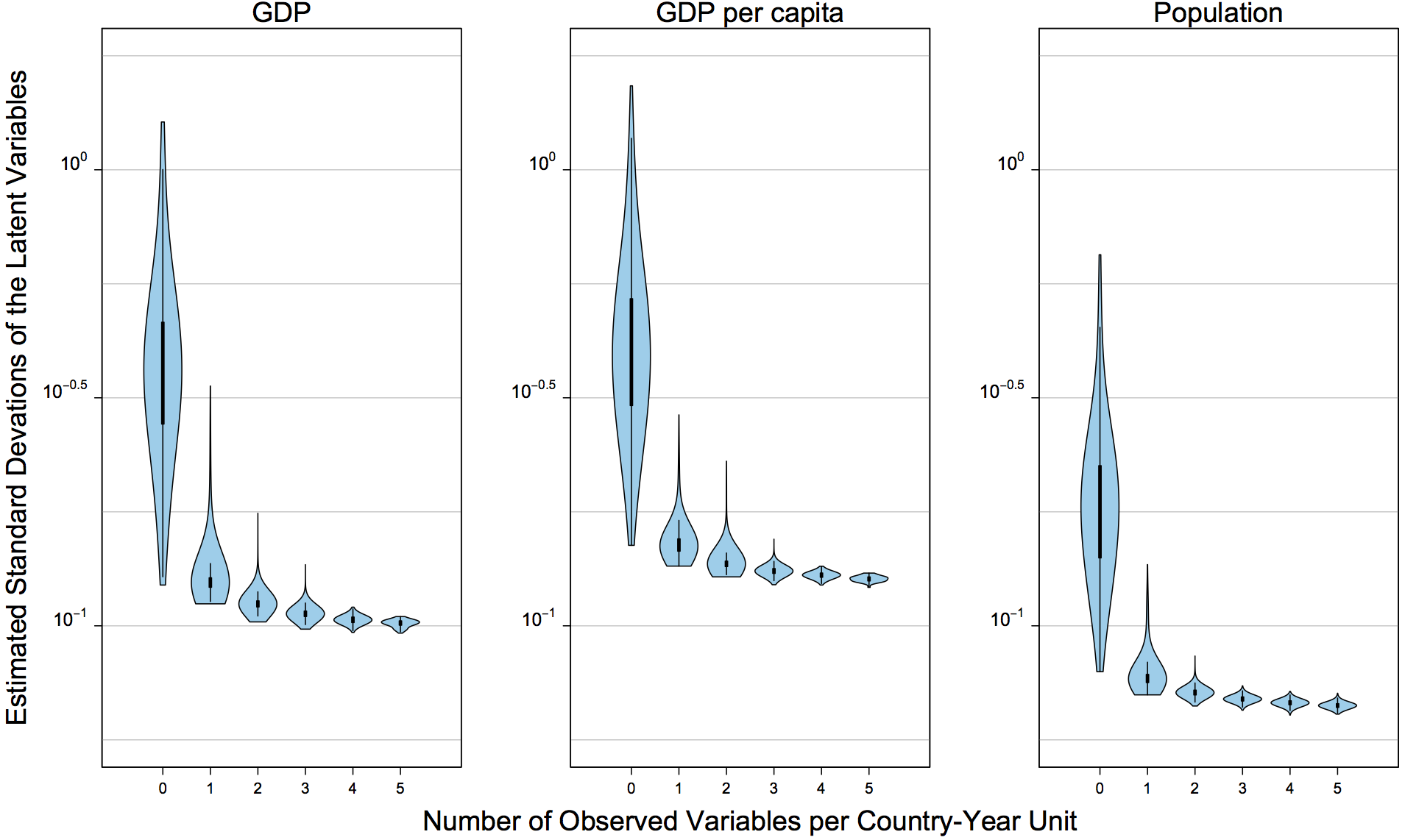}
\caption{The plot illustrates the relationship between the amount of information --- the number of observed variables per country-year unit --- and the level of uncertainty for each country-year estimate of the latent variable. As the number of observed pieces of information increases, the level of uncertainty decreases. Because the latent variable model is dynamic, even country-units with 0 observed variables have some information --- the latent variable estimate from the prior year --- with which to estimate a value of the latent variable. For this reason, the amount of variation of the latent standard deviation is greatest for these units without information. The further away a unit exists to an observed variable, the greater the uncertainty and the estimate of the standard deviation. Importantly, the level of uncertainty is a country-year unit parameter. This parameter can be used to incorporate uncertainty in empirical models that include the new estimates of GDP, GDP per capita, and population. See for example \citet{SchnakenbergFariss:2014}.}
\label{Latent_SD_Rplot.png}
\end{center}
\end{figure}

Theoretically, the precision of our posterior estimates should increase with the amount of information (i.e. number of component measures) available for each country year. Figure \ref{Latent_SD_Rplot.png} illustrates the relationship between the amount of information --- the number of observed variables per country-year unit --- and the level of uncertainty for each country-year estimate of the latent variable. As the number of observed pieces of information increases, the level of uncertainty decreases. Because the latent variable model is dynamic, even country-year units with 0 observed variables have some information --- the latent variable estimate from the prior year --- with which to estimate a value of the latent variable. We use these intervals to estimate the precision of the new posterior predictions relative to the original observed variable.

To assess the precision of the new posterior predictions relative to the original observed variable we estimate a country-year unit Z-score. Because the original variables and the predicted variables are both continuous, interval-level, and normally distributed, we are able to compare the original observed country-year unit variable's position relative to the country-year unit distribution of the posterior predicted interval. The Z-score's value represents the positions of the observed variable's value relative to the center of the posterior predicted interval. A value of 0 indicates that the observed variable's value falls directly at the center of the posterior predicted interval. Z-score units above and below the 0 represent standard deviation differences from the center value of the posterior predicted interval. We can therefore use these values to get a sense of how close the observed value for each country-year variable is to its corresponding posterior predicted interval. The country-year-variable Z-score values take the form: $z_{itj} = \frac{y_{itj} - E(\tilde{y}_{itj})}{\sigma_{\tilde{y}_{itj}}} $, where $y_{itj}$ is the observed value for the country-year-variable,  $E(\tilde{y}_{itj})$ is the expected value or mean for the posterior predicted interval, and $\sigma_{\tilde{y}_{itj}}$ is the standard deviation for the posterior predicted interval.

\begin{figure}[ht]
\begin{center}$
\begin{array}{c}
\includegraphics[width=6.0in]{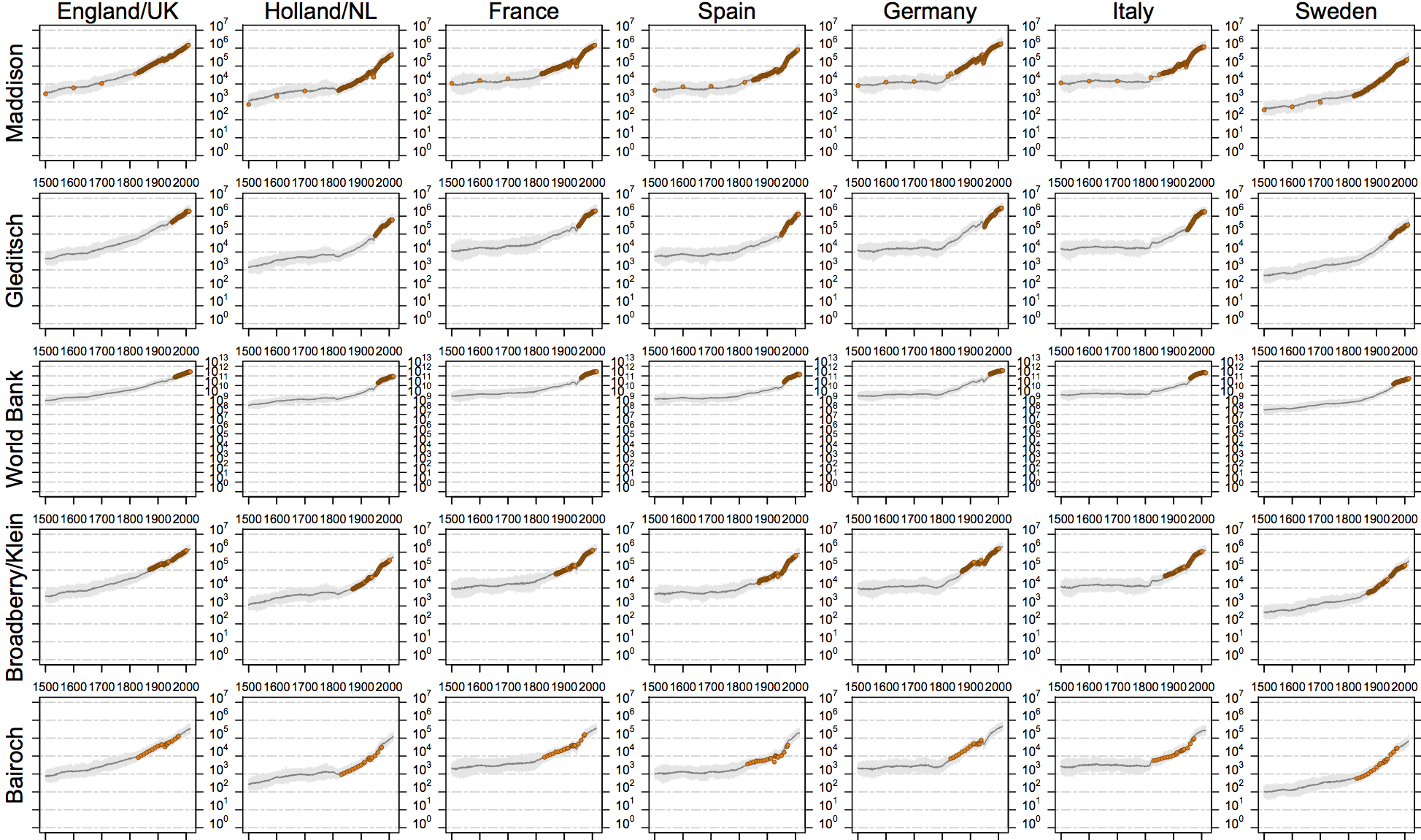} \\
\\
\includegraphics[width=6.0in]{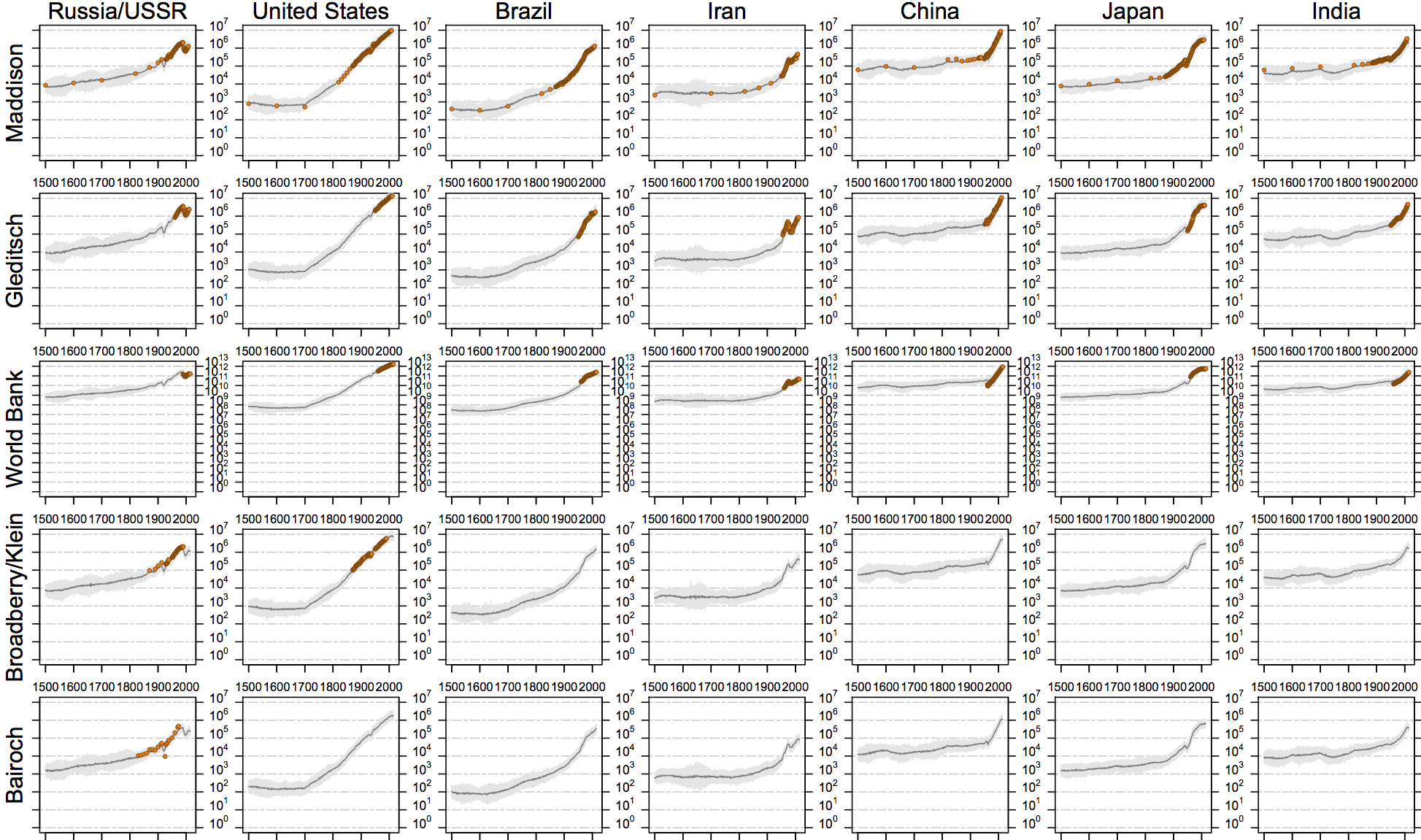} \\
\end{array}$
\end{center}
\caption{Gross Domestic Product posterior prediction intervals (grey lines) and observed variables (orange points). Approximately 86 percent of the observed country-year-variable values fall with $\pm$ 1 standard deviation of the posterior predicted range of estimates generated from the model.}
\label{GDP_estimates}
\end{figure}

\begin{figure}[ht]
\begin{center}$
\begin{array}{c}
\includegraphics[width=6.0in]{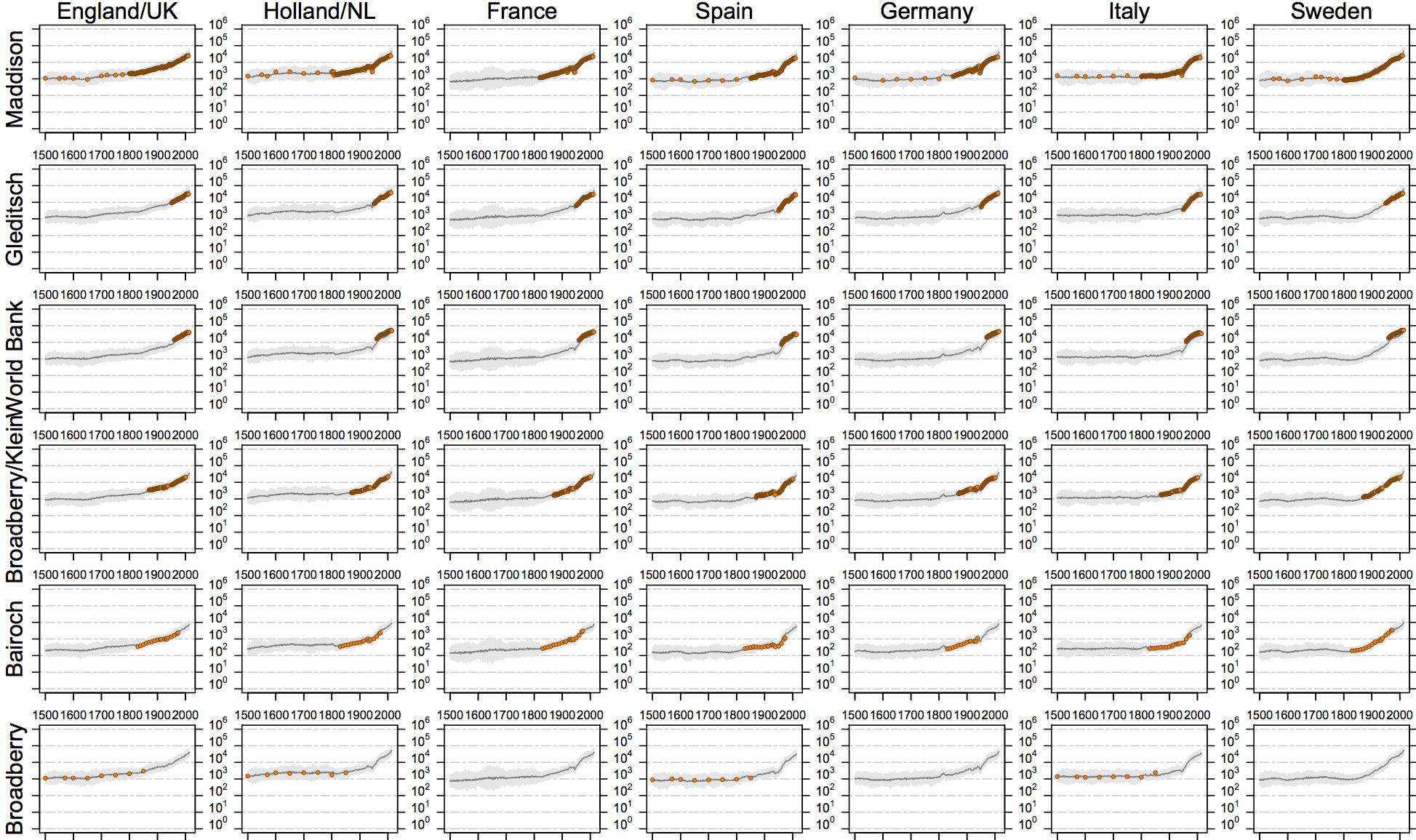} \\
\\
\includegraphics[width=6.0in]{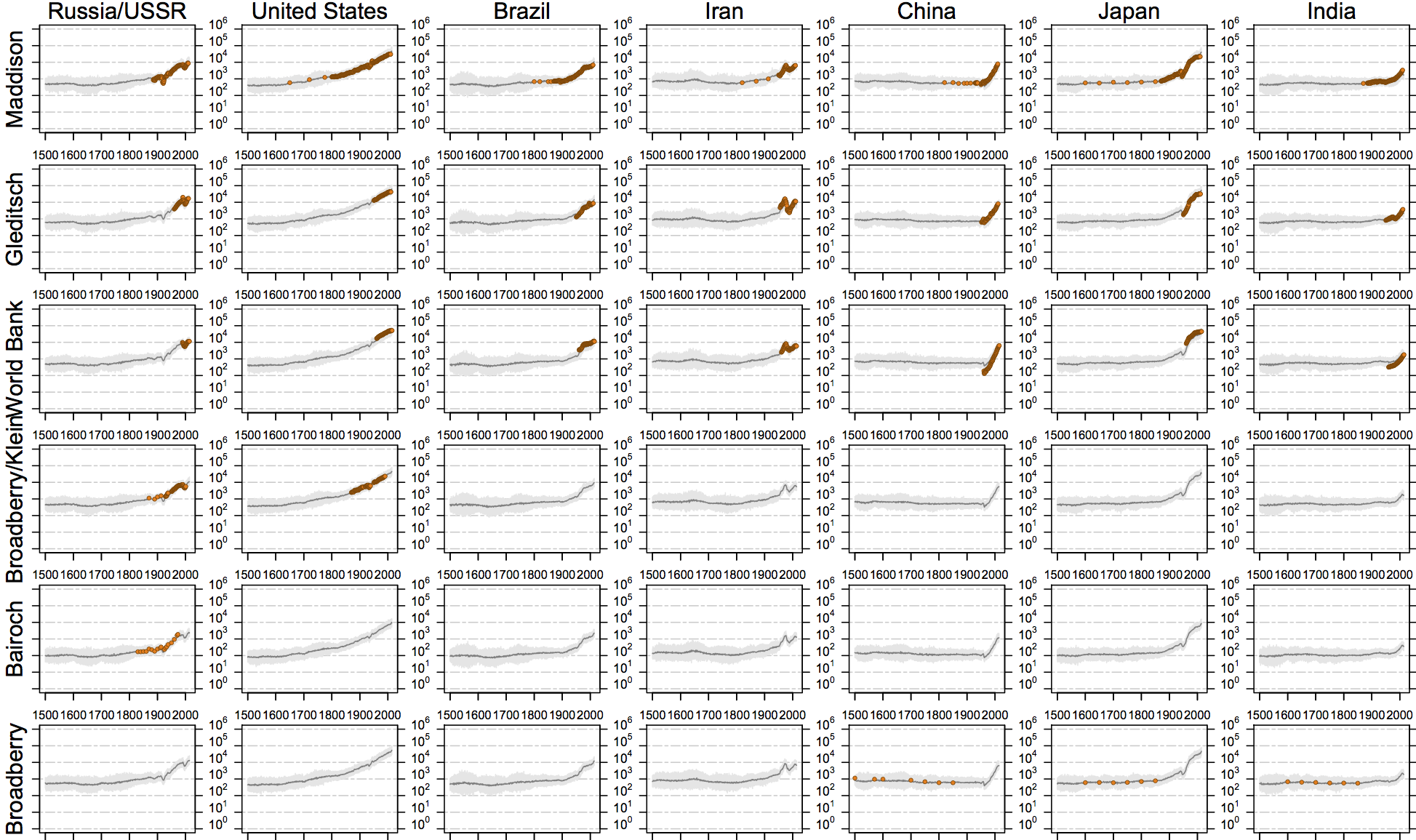} \\
\end{array}$
\end{center}
\caption{Gross Domestic Product per capita posterior prediction intervals (grey lines) and observed variables (orange points). Approximately 86 percent of the observed country-year-variable values fall with $\pm$ 1 standard deviation of the posterior predicted range of estimates generated from the model.}
\label{GDPPC_estimates}
\end{figure}

\begin{figure}[ht]
\begin{center}$
\begin{array}{c}
\includegraphics[width=6.0in]{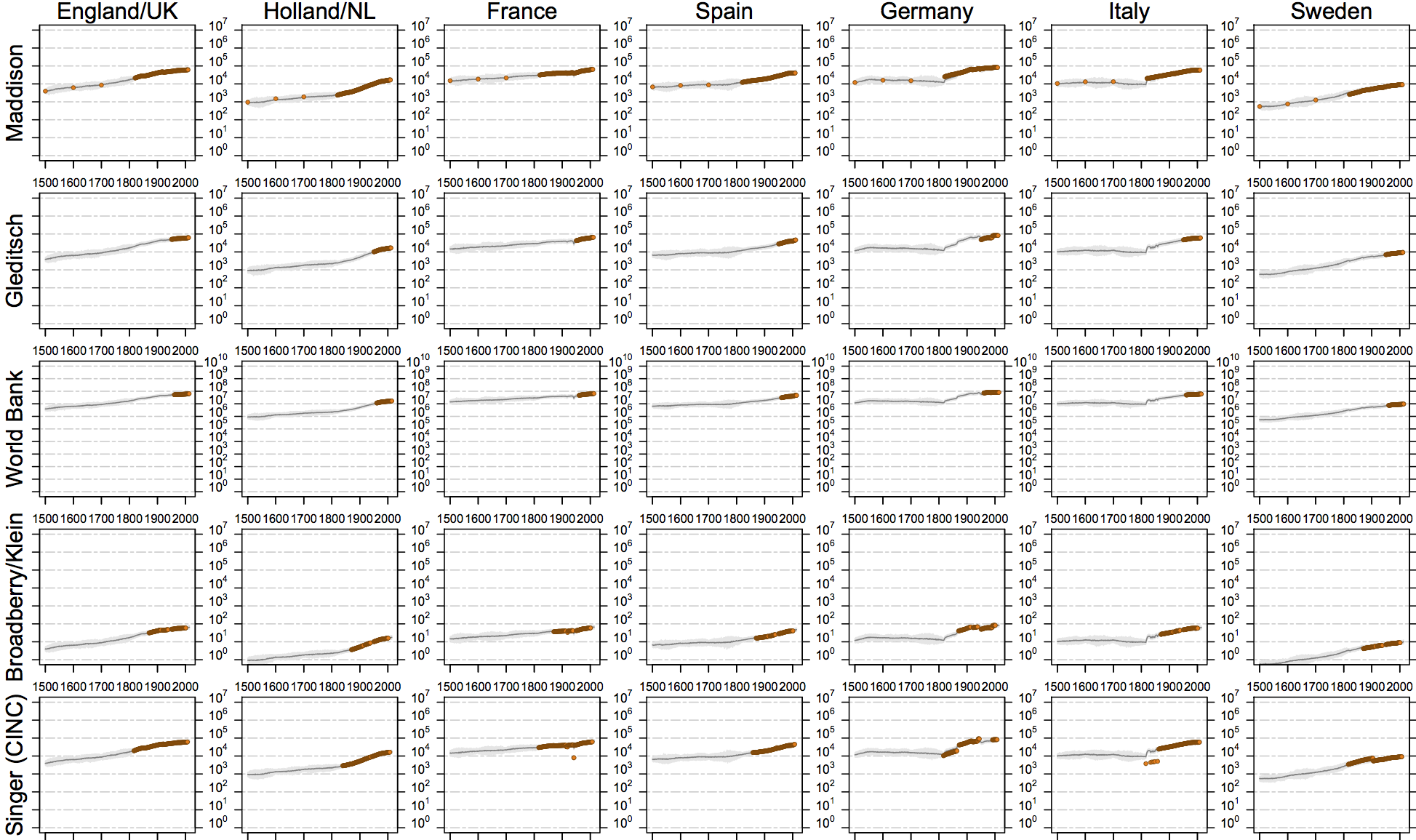} \\
\\
\includegraphics[width=6.0in]{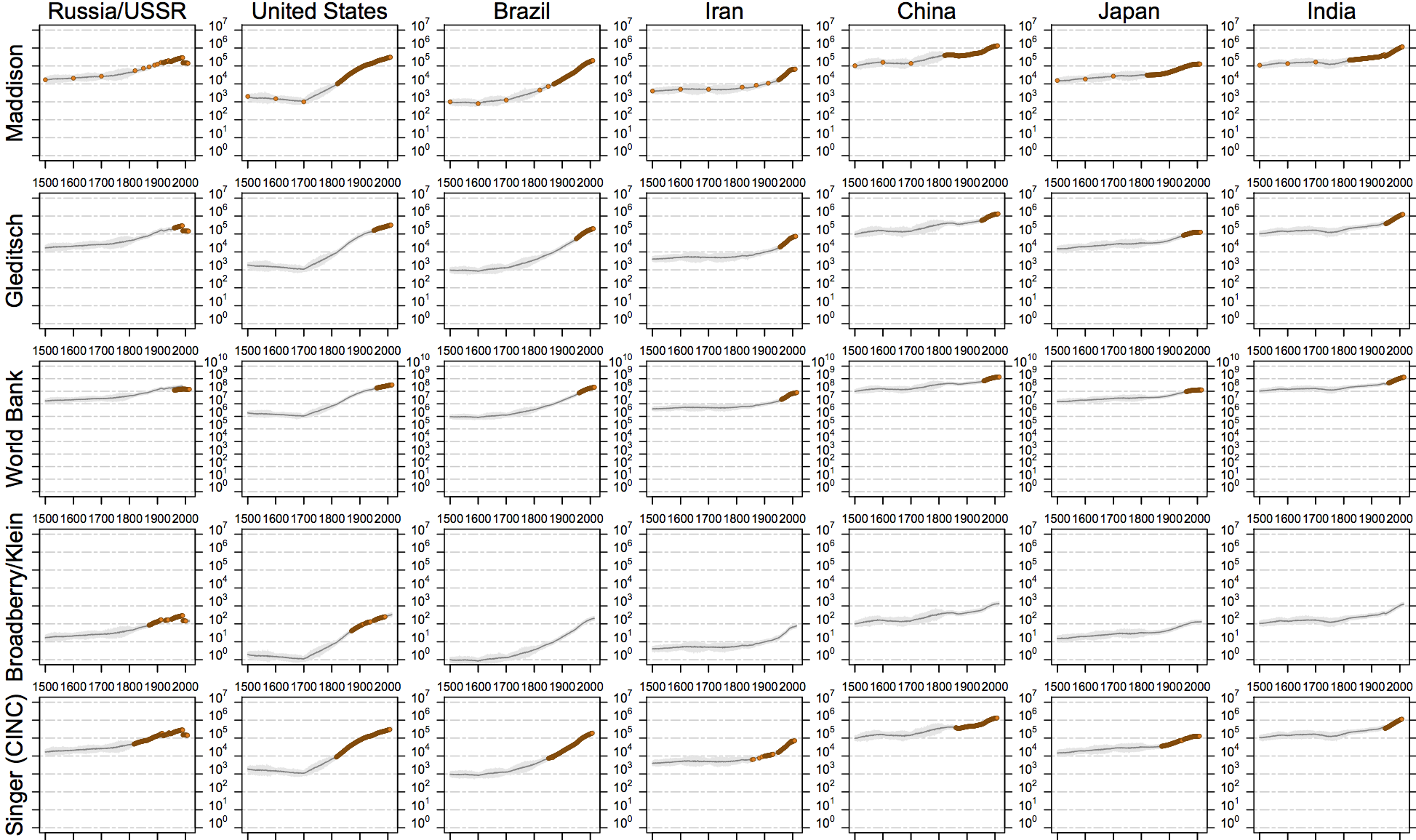} \\
\end{array}$
\end{center}
\caption{Population posterior prediction intervals (grey lines) and observed variables (orange points). Approximately 86 percent of the observed country-year-variable values fall with $\pm$ 1 standard deviation of the posterior predicted range of estimates generated from the model.}
\label{POP_estimates}
\end{figure}

Table \ref{Proportion_Table} quantifies the similarity between the original data and the full range of possible values for the posterior predicted range using the country-year-variable Z-scores $z_{itj}$.  Approximately 86 percent of the observed country-year-variable values are within $\pm$ 1 standard deviation of the predicted range of the estimated data. We can visualize these patterns as well. The level of uncertainty over time is visible in time series plots for each of the new posterior predictions for 14 different countries (see Figure \ref{GDP_estimates}, \ref{GDPPC_estimates}, and \ref{POP_estimates}). These figures illustrate the high level of agreement between the original observed variables and the new posterior prediction intervals. These intervals are much larger for country-year units without data as well as country-year units with fewer observed values. Overall, the new estimates closely approximate the original data, but do not always agree. These areas of disagreement between original observed variable and posterior prediction represent useful deviant cases that can inform future research \citep[e.g.,][]{Lijphart:1971, SeawrightGerring:2008}.

\clearpage

\begin{table}[ht]
\centering
\small
\begin{tabular}{| l | ccc |r |}
 \multicolumn{5}{c}{Proportion of Observed Values within Model Prediction Interval}   \\
  \hline
 & $\pm 1 \sigma$ & $\pm 2 \sigma$ & $\pm 3 \sigma$ & $n$ \\
  \hline
Maddison GDP 	&	0.8430	&	0.9621	&	0.9873	&	11794	\\
 Gleditsch GDP 	&	0.8187	&	0.9566	&	0.9870	&	8724	\\
 World Bank GDP 	&	0.6343	&	0.9264	&	0.9920	&	8107	\\
Broadberry \&  Klein GDP  	&	0.9542	&	0.9812	&	0.9996	&	2338	\\
Bairoch GNP	&	0.8217	&	0.9518	&	0.9831	&	415	\\
 \hline
  Maddison GDPPC	&	0.8484	&	0.9666	&	0.9897	&	13068	\\
  Gleditsch GDPPC 	&	0.8152	&	0.9574	&	0.9872	&	8724	\\
  World Bank GDPPC 	&	0.6178	&	0.9253	&	0.9906	&	7970	\\
  Bairoch GDPPC 	&	0.9567	&	0.9965	&	1.0000	&	2566	\\
  Broadberry GDPPC 	&	0.8645	&	0.9837	&	1.0000	&	369	\\
  Broadberry \& Klein GDPPC	&	0.9412	&	1.0000	&	1.0000	&	51	\\
 \hline
 Maddison pop 	&	0.9512	&	0.9790	&	0.9871	&	14138	\\
 Gleditsch pop 	&	0.9583	&	0.9858	&	0.9897	&	8724	\\
 World Bank pop 	&	0.9606	&	0.9856	&	0.9886	&	10646	\\
 Broadberry \	&	0.9400	&	0.9725	&	0.9757	&	2802	\\
 Singer (CINC) pop 	&	0.9372	&	0.9729	&	0.9817	&	12282	\\
  \hline
  Weighted Proportion &	0.8599	&	0.9655	&	0.9881	&		\\
   \hline
\end{tabular}
\caption{Proportion of the observed variable values that are within the estimated prediction interval for each of the estimated variables. For example, on average, approximately 86 percent of the observed country-year-variable values fall with $\pm$ 1 standard deviation of the posterior predicted range of estimates generated from the model.  Overall, the model is able to estimate the original observed values quite well even while taking into account information across all three types of variables.}
\label{Proportion_Table}
\end{table}

%\begin{figure}[ht]
%\begin{center}$
%\begin{array}{cc}
%\includegraphics[width=3.250in]{plots/GDP_estimates_1.png} &
%\includegraphics[width=3.250in]{plots/GDP_estimates_2.png} \\
%\includegraphics[width=3.250in]{plots/GDPPC_estimates_1.png} &
%\includegraphics[width=3.250in]{plots/GDPPC_estimates_2.png} \\
%\includegraphics[width=3.250in]{plots/POP_estimates_1.png} &
%\includegraphics[width=3.250in]{plots/POP_estimates_2.png} \\
%\end{array}$
%\end{center}
%\caption{GDP, GDP per capita, and Population posterior prediction intervals (grey lines) and observed variables (orange points) for 14 country series. Approximately 86 percent of the observed country-year-variable values fall with $\pm$ 1 standard deviation of the posterior predicted range of estimates generated from the model.}
%\label{ALL_estimates}
%\end{figure}

\clearpage
\subsection*{Identifying Bias in Component Measures}
We have already demonstrated that the number of observed variables is related to the level of uncertainty for each country-year estimate of the latent variable and the level of uncertainty around the posterior prediction intervals (see Figure \ref{Latent_SD_Rplot.png}).  We also illustrated how this manifests using the Z-score estimates $z_{itj}$ to show how closely the prediction interval is to the observed value for each country-year-variable observation. However, as additional variables are observed for a given country-year unit we begin to observe bias with respect to the distance between the observed value of a given variable and the posterior prediction interval as evidenced by the value of $z_{itj}$. These statistics allow us to gain insight into how the latent variable model accounts for disagreement between the observed variables.

\begin{figure}[ht]
\begin{center}$
\begin{array}{c}
\includegraphics[width=4.0in]{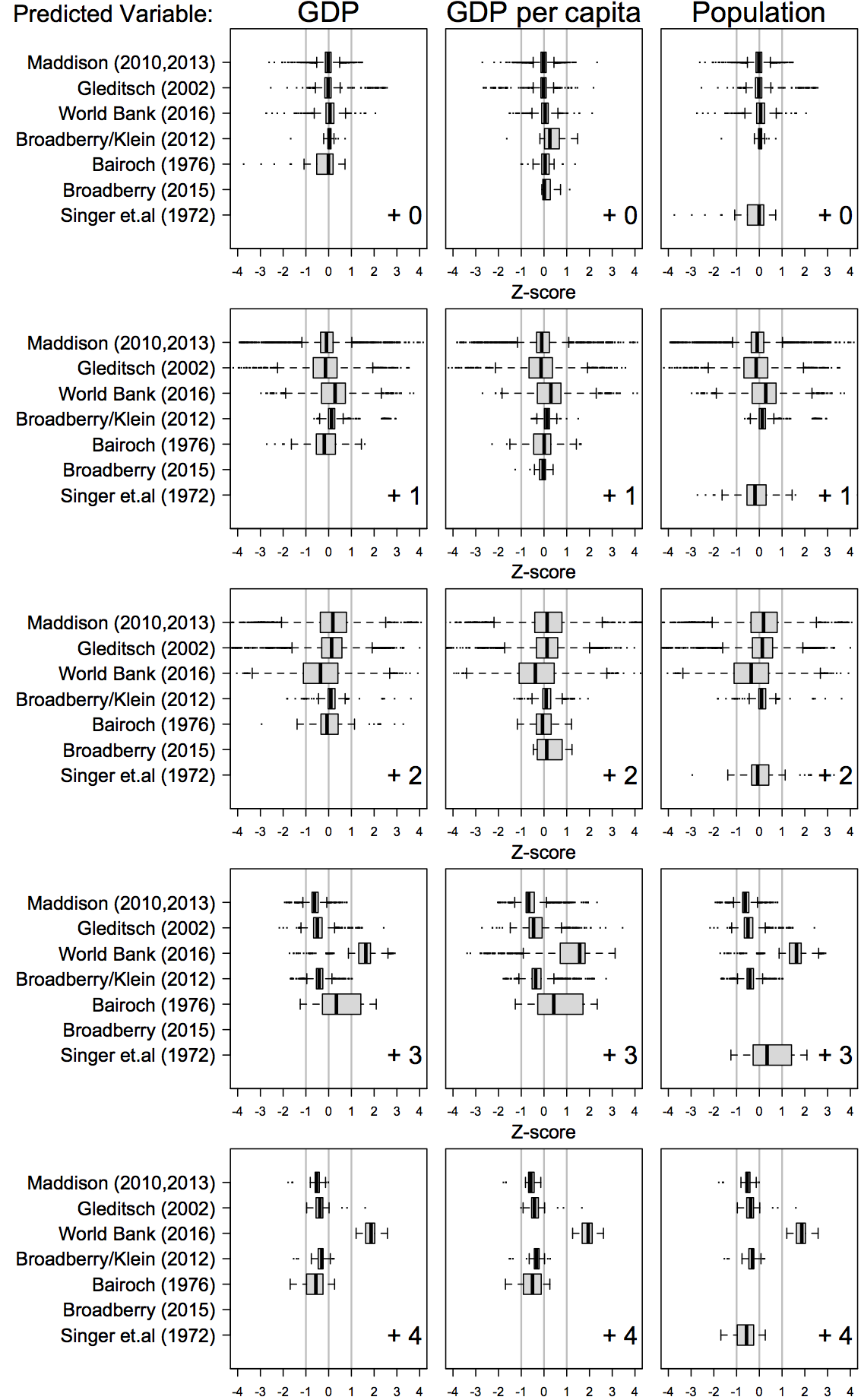}
\end{array}$
\end{center}
\caption{This figure illustrates the relationship between the number of observed variables and the level agreement between the original country-year variable value and the posterior predicted intervals for country-year units for which we observed the labeled variable in addition to 0 to 4 of the additional variables. The five rows of panels represent the country-year units with 1 to 5 observed variables (0 to 4 additional variables).  For a small number of country-year units that are covered by 4 or 5 of the available variables, we begin to see some substantial differentiation between the prediction intervals and the observed values.}
\label{item_bias.png}
\end{figure}

Figure \ref{item_bias.png} illustrates the relationship between the number of observed variables and the level of agreement between the original country-year variable value and the posterior predicted intervals for country-year units for which we observed the labeled variable in addition to $+$0 to $+$4 of the additional variables. The five rows of panels represent the country-year units with 1 to 5 observed variables ($+$0 to $+$4 additional observed variables).  When only one of the original variables is observed for a given country-year unit, we see that both the level of uncertainty is relatively high (see Figure \ref{Latent_SD_Rplot.png}) and the distance between the observed value and prediction interval is relatively close (see Figure \ref{item_bias.png}). That is, for country-year units with only 1 observed variable, the Z-scores are centered over 0 and the standard deviations for the posterior predicted values are relatively high. These two patterns change as additional variables are observed for a given country-year unit. For the small number of country-year units that are covered by 4 or 5 of the available variables, we begin to see some substantial differentiation between the prediction intervals and the observed values. The values for $z_{itj}$ are nearly all different than 0 for these country-year units. This is especially apparent for the observed values from the World Bank variables. According to the model based estimates, the original observed values from the World Bank are systematically higher than the posterior prediction intervals, though the observed values still tend to fall within 2 standard deviations of these intervals. The values from the variables from the other datasets tend to fall below the median prediction but are within 1 standard deviation of the intervals.

Overall, as more information becomes available for a particular country-year unit, the latent variable model becomes increasingly certain about the value of the estimated posterior prediction interval ($\sigma_{\tilde{y}_{itj}}$ decreases). However, as the number of observed values increases we can begin to identify systematic differences between the original data and our new posterior predicted estimates. These country-year units in particular are useful cases to study because they may represent places where information is collected in different ways or where other data production biases might be introduced into some of the constituent datasets we have brought together. The underlying reasons for these systematic differences are useful starting points for future scholarship. One advantage of our model, is that it identifies these areas when there is sufficient coverage across available datasets.

\clearpage
\subsection*{Improving Model Fit with Knowledge of the Original Data}
With knowledge about the origin of the original observed data, we can assess the value added of our model compared to some other methods used to generate estimates for units that are missing observations. Several of the observed datasets \citep{gleditsch2002, singeretal1972, wb2016} provide information about the methods used to infer missing values using interpolation or extrapolation. In all of the models we have discussed so far, we have removed these interpolated or extrapolated values in order to let our model predict them. Table \ref{DIFF_TAB} displays the difference in root-mean-square error (RMSE) comparing model predictions for several of these variables between two versions of the latent variable model. The models are identical except that in the primary model, units with interpolated or extrapolated values are changed to missing for each variable when those units are identified within the source material that accompanies the original dataset. Smaller RMSE statistics indicate that, when the interpolated or extrapolated values are present, that they increase the uncertainty in the estimates for those units that were observed from primary or secondary sources materials (i.e., those values that were not interpolated or extrapolated). 

Only the five variables included in this table have accompanying information, which indicates which units are not drawn from primary sources documents and are estimated instead. Overall, the evidence indicates that for three of these five variables (all of the population variables), the latent variable model predictions achieve better model fit when the interpolated and extrapolated values are set to missing. Additional auxiliary information about the data sources and potential biases of specific country-year units might help to further improve the model-based estimates. The model we have developed is extendable and can accommodate such information as it becomes available.

\begin{table}[htbp!]
\centering
\small
\begin{tabular}{| l | rrr |}
  \hline
 & $Diff$  & $95\%~Credible~Int$  &  $Pr(Diff)$ \\
  \hline
World Bank pop & -0.015 & [-0.018, -0.013] & 1.000 \\
Gleditsch GDP & 0.001 & [-0.007, ~0.009] & 0.363 \\
Gleditsch GDPPC & 0.001 & [-0.006, ~0.008] & 0.374 \\
Gleditsch pop & -0.013 & [-0.016, -0.010] & 1.000 \\
Singer (CINC) pop & -0.006 & [-0.009, -0.004] & 1.000 \\
   \hline
\end{tabular}
\caption{Difference in root-mean-square error (RMSE) comparing model predictions for several variables between two versions of the latent variable model. The models are identical except in one model, units with interpolated values are changed to missing for each variable when those units are identified within the source material that accompanies the original dataset. Smaller RMSE statistics indicate that the interpolated or extrapolated values are increasing the uncertainty in the estimates of the units without interpolated or extrapolated values. Only the five variables included in this table have accompanying information, which indicates which units are not drawn from primary sources documents but are estimated instead. Overall, the evidence indicates that for three of these five variables (all of the population variables), the latent variable model predictions achieve better model fit when the interpolated and extrapolated values are set to missing.}
\label{DIFF_TAB}
\end{table}

\clearpage
\subsection*{Extending the Model with Unit Specific Information: The Case of Ghana}
\citet{Jerven:2014}  provides new estimates of GDP growth in the Gold Coast and Ghana using sources from the British colonial administration. The information is historically of interest because of the large number of primary source documents that account for the period before, during, and after a major cocoa boom, which is not picked up by available data. This extension illustrates how a country-specific analysis can help to augment the latent variable model that we described above. 

\begin{figure}[ht]
\begin{center}$
\begin{array}{c}
\includegraphics[width=6.0in]{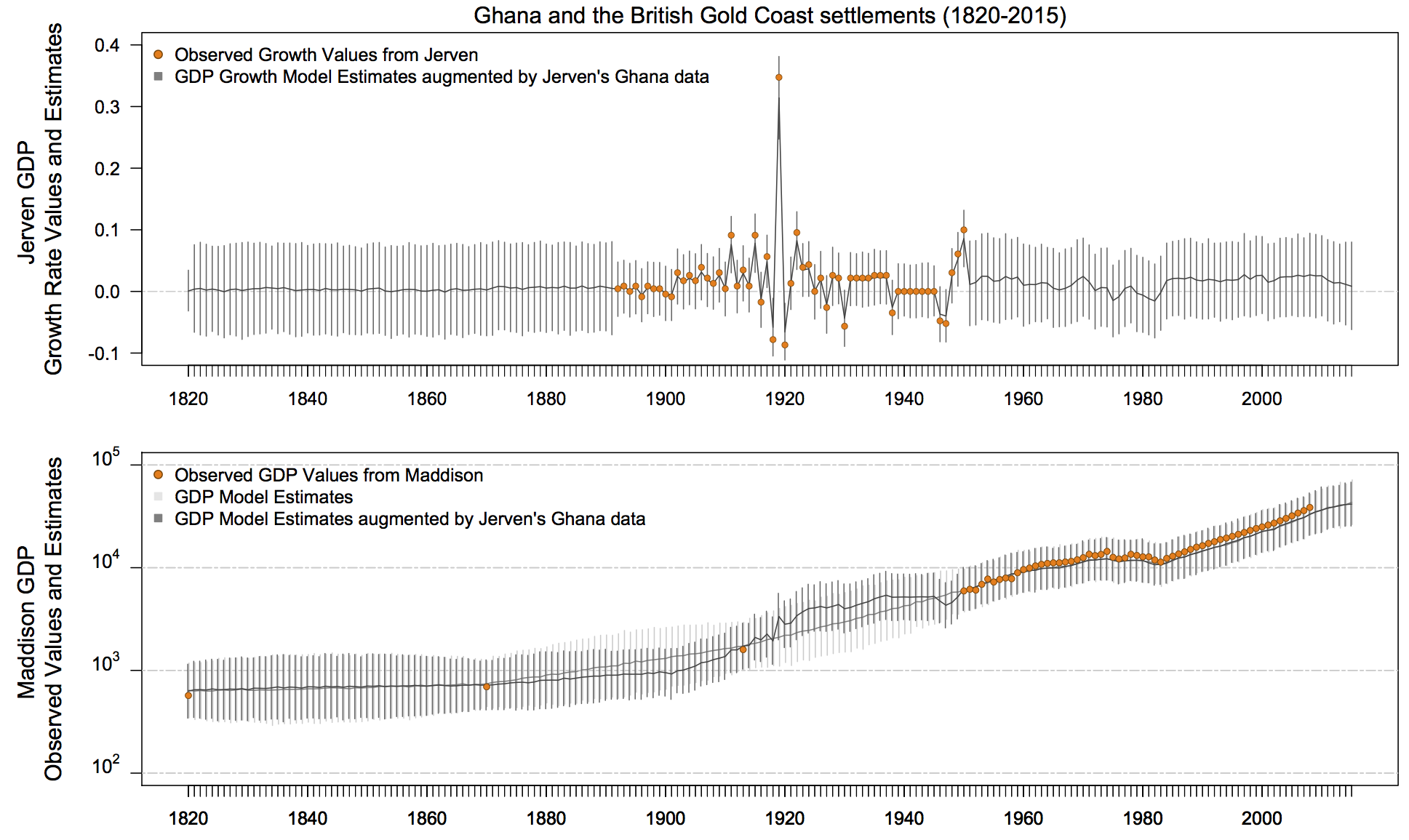}
\end{array}$
\end{center}
\caption{Model extension for Ghana using country specific information collected from country primary and secondary source material by \citet{Jerven:2014}. From 1891 through the 1950s, the economy in Ghana grew rapidly before eventually slowing because of the boom in cocoa production. These changes are now captured in the time series for Ghana and replicate the point estimates produced by \citet{Jerven:2014}. }
\label{Ghana.png}
\end{figure}

Because the growth rate is a deterministic transformation of GDP, we can include observed information on this concept in the model developed above. To extend the model we add a fourth category $k: \Delta gdp$ and a new set of observed indicators for this new category such that  $\pi_{\Delta gdp} = \{ y_{it17})$. $y_{it17}$ is the growth rate variable for Ghana taken from \citet{Jerven:2014}. We set $\alpha_{17}$ to 0 because the growth rate variable is very close to centered around 0 and we do not need to scale it. Just like the GDP per capita latent variable, the new growth rate latent variable is a deterministic transformation of other latent variables in the model. Specifically, we link these new growth rate data to the latent GDP variable in years $t$ and $t-1$, using the form $\theta_{it, \Delta gdp} \leftarrow \frac{\theta_{it, gdp}}{\theta_{it-1, gdp}}-1$. The likelihood equation remains unchanged with these additions to the latent variable model.\protect\footnote{We also estimate an additional $\tau_{\Delta gdp}$ parameter.} With this expanded parameterization, we can now leverage the growth rate data collected by \citet{Jerven:2014} to improve the estimation of the GDP, and GDP per capita latent variables for this country. Figure \ref{Ghana.png} shows the improvement in the Ghana specific GDP time series. From 1891 through the 1950s, the economy in Ghana fluctuated and grew rapidly before eventually slowing after the boom in cocoa production. These historic changes are now captured in the time series for Ghana and replicate the point estimates produced by \citet{Jerven:2014}. We can now illustrate these changes directly and how they change the GDP trend over time. 

\clearpage

\section*{Conclusion}
In this paper, we identified and discussed the limitations of existing GDP, GDP per capita, and population variables. We then introduced and described a new latent variable model that estimates posterior predictions for these important concepts. These new estimates provide several primary advantages over existing measures: (1) they extend the temporal and spatial coverage of existing measures, (2) they include multiple manifest indicators of each concept, (3) they quantify the uncertainty for each of the country-year estimates. In addition to the new estimates, the latent variable model itself provides a principled avenue for exploring the relationship between new data and existing data. To close, we outline several possible extensions of our latent variable model. To encourage the extension of our model, we make publicly available the data and code used to construct it. The code in our replication archive has been carefully annotated to encourage easy replication, modification, and we hope theoretically informed extensions.

One possible modification of the latent variable is the inclusion of additional manifest variables. Though we have included many sources of data for GDP, GDP per capita, and population there are other sources that we have not yet explored. Researchers, could extend our model by adding alternative population measures, such as those from the World Prospects Series and the International Programs Center, or alternative GDP measures from regional organizations, such as the OECD \citep{oecd2016}. Additional data could also be collected on when a country's base year is updated or when a country adheres to the recommended system of national accounting. Data from country specific studies could also be added to the model as the case of Ghana illustrates \citep{Jerven:2014}. 

Another possible modification is to explore the inclusion of manifest variables plausibly related to the reliability of state and third-party estimates of GDP, GDP per capita, and population information. These covariates could include measures of state capacity such as bureaucratic quality or tax capacity \citep{hendrix2010}, the size of the informal economy \citep{ifc2015}, or even survey-based data such as the level of perceived corruption \citep{ti2015,kaufmannetal2010}. Though each of these sources of data likely have their own measurement issues, leveraging an understanding of the different biases inherent to different data sources can help improve the validity of the resulting estimates \cite[e.g.,][]{Coppedgeetal:2014, Fariss:2014, Fariss:2015:treaty, PemsteinMarquardtTzelgovWangMiri:2015, PemsteinMeserveMelton:2010, PemsteinTzelgovWang:2015, SchnakenbergFariss:2014}. Overall, we believe that the new latent variables estimates of GDP, GDP per capita, and population, now available for over 500 years of history, will be of use to scholars interested in political and economic theories of interstate and intrastate behaviors.

\clearpage

\singlespacing
\bibliographystyle{chicago}
\bibliography{bib_20111208,main,historical}

\clearpage
\section*{Appendix}
% GNP vs GDP
%\vspace{5mm}
\subsubsection*{Yearly Correlation of GNP compared to GDP measures}
% latex table generated in R 3.3.1 by xtable 1.8-2 package
% Mon Dec 12 07:44:34 2016
\begin{table}[ht]
\centering
\begingroup\footnotesize
\begin{tabular}{rrllll}
  \toprule
Year & N (Bairoch) & Broadberry/Klein (2012) & Gleditsch (2002) & Maddison (2010) & World Bank (2016) \\ 
  \midrule
1830 &  15 &  &  & 0.89 [0.55, 0.98] &  \\ 
  1840 &  16 &  &  & 0.87 [0.36, 0.98] &  \\ 
  1850 &  17 &  &  & 0.91 [0.75, 0.97] &  \\ 
  1860 &  19 &  &  & 0.92 [0.73, 0.98] &  \\ 
  1870 &  19 & 0.99 [0.98, 1.00] &  & 0.92 [0.79, 0.97] &  \\ 
  1880 &  19 & 0.99 [0.98, 1.00] &  & 0.93 [0.78, 0.98] &  \\ 
  1890 &  19 & 1.00 [0.99, 1.00] &  & 0.90 [0.75, 0.96] &  \\ 
  1900 &  19 & 1.00 [0.99, 1.00] &  & 0.92 [0.80, 0.97] &  \\ 
  1910 &  19 & 0.99 [0.98, 1.00] &  & 0.89 [0.72, 0.96] &  \\ 
  1913 &  19 & 0.99 [0.98, 1.00] &  & 0.91 [0.79, 0.97] &  \\ 
  1925 &  23 & 0.85 [0.63, 0.94] &  & 0.86 [0.68, 0.95] &  \\ 
  1928 &  23 & 0.99 [0.98, 1.00] &  & 0.99 [0.97, 1.00] &  \\ 
  1929 &  23 & 0.99 [0.97, 1.00] &  & 0.99 [0.97, 0.99] &  \\ 
  1933 &  23 & 0.99 [0.98, 1.00] &  & 0.99 [0.97, 0.99] &  \\ 
  1937 &  23 & 0.99 [0.98, 1.00] &  & 0.99 [0.97, 1.00] &  \\ 
  1938 &  23 & 0.99 [0.98, 1.00] &  & 0.99 [0.97, 1.00] &  \\ 
  1950 &  24 & 0.99 [0.97, 0.99] & 0.98 [0.93, 0.99] & 0.99 [0.97, 1.00] &  \\ 
  1960 &  24 & 0.98 [0.95, 0.99] & 0.91 [0.78, 0.97] & 0.98 [0.96, 0.99] & 0.98 [0.93, 0.99] \\ 
  1970 &  24 & 0.96 [0.92, 0.99] & 0.90 [0.78, 0.96] & 0.97 [0.92, 0.99] & 0.98 [0.95, 1.00] \\ 
  1973 &  24 & 0.96 [0.90, 0.98] & 0.90 [0.78, 0.96] & 0.96 [0.91, 0.99] & 0.99 [0.95, 1.00] \\ 
   \bottomrule
\end{tabular}
\endgroup
\caption{The table presents the correlation coefficient between the GNP estimates from Bairoch (1976) and alternative GDP measures. The 95\% confidence intervals is indicated in square brackets. All correlation coefficients are statistically significant at the minimum 5\% level. The correlation between GNP and GDP is strong with correlation coefficients between 0.865 and 0.996 for the period 1830-1973. The second column indicates the maximum number of observations that the correlation coefficient is based on.} 
\label{gnp_gdp}
\end{table}

\clearpage
\subsubsection*{Summary Statistics of Observed Variables}

% Table created by stargazer v.5.2 by Marek Hlavac, Harvard University. E-mail: hlavac at fas.harvard.edu
% Date and time: Mon, Dec 12, 2016 - 14:25:49
\begin{table}[!htbp] \centering 
  \caption{Summary statistics for components of latent variable models.} 
  \label{} 
\footnotesize 
\begin{tabular}{@{\extracolsep{5pt}}lcccccc} 
\\[-1.8ex]\hline 
\hline \\[-1.8ex] 
Statistic & \multicolumn{1}{c}{Min} & \multicolumn{1}{c}{Median} & \multicolumn{1}{c}{Mean} & \multicolumn{1}{c}{Max} & \multicolumn{1}{c}{St. Dev.} & \multicolumn{1}{c}{N} \\ 
\hline \\[-1.8ex] 
GDP Maddison (2010): 1500-2008 & 1.099 & 9.647 & 9.760 & 16.065 & 1.950 & 11,794 \\ 
GDP Gleditsch (2002): 1950-2011 & 2.944 & 9.900 & 9.947 & 16.395 & 2.307 & 8,724 \\ 
GDP World Bank (2016): 1960-2015 & 16.881 & 23.405 & 23.616 & 30.437 & 2.427 & 8,107 \\ 
GDP Broadberry/Klein (2012): 1870-2001 & 7.114 & 11.138 & 11.068 & 15.574 & 1.569 & 2,338 \\ 
GNP Bairoch (1972): 1830-1973 & 5.298 & 8.691 & 8.698 & 13.063 & 1.467 & 415 \\ 
GDP p.c. Maddison (2013): 1500-2000 & 5.315 & 7.665 & 7.762 & 10.667 & 1.002 & 13,068 \\ 
GDP p.c. Gleditsch (2002): 1950-2011 & 4.889 & 8.319 & 8.399 & 13.357 & 1.252 & 8,724 \\ 
GDP p.c. World Bank (2016): 1960-2015 & 4.749 & 8.069 & 8.151 & 11.886 & 1.498 & 7,970 \\ 
GDP p.c. Broadberry/Klein (2012): 1870-2001 & 6.696 & 8.390 & 8.446 & 10.154 & 0.793 & 2,566 \\ 
GNP p.c. Bairoch (1976): 1830-1973 & 5.136 & 6.242 & 6.347 & 8.159 & 0.727 & 369 \\ 
GDP p.c. Broadberry (2015): 1500-1850 & 6.321 & 7.012 & 6.991 & 8.005 & 0.474 & 51 \\ 
Population Maddison (2010): 1500-2010 & 3.223 & 8.749 & 8.872 & 14.102 & 1.619 & 14,138 \\ 
Population Gleditsch (2002): 1950-2011 & 1.792 & 8.713 & 8.455 & 14.096 & 2.177 & 8,724 \\ 
Population World Bank (2016): 1960-2013 & 8.397 & 15.390 & 15.101 & 21.039 & 2.230 & 10,646 \\ 
Population Broadberry/Klein (2012): 1870-2001 & 0.207 & 2.288 & 2.541 & 5.665 & 1.219 & 2,802 \\ 
Population Singer et al. (1972): 1816-2001 & 2.197 & 8.761 & 8.713 & 14.097 & 1.801 & 12,285 \\ 
\hline \\[-1.8ex] 
\multicolumn{7}{l}{The exact measurement of the GDP, per capita GDP, and population variables differs across different sources.} \\ 
\multicolumn{7}{l}{See variable description tables for details on the measurement.} \\ 
\end{tabular} 
\end{table}

\clearpage
\subsubsection*{JAGS Model Code}
\singlespace
{\scriptsize
\begin{Verbatim}[frame=single]
model{
 for(i in 1:n){
   # GDP items
   for(j in 1:5){
     xb[i,j] <- b0[j] + b1[j] * x.gdp[i]
     y[i,j] ~ dnorm(xb[i,j], tau1)
     y.star[i,j] ~ dnorm(xb[i,j], tau1)
    }

   # Population items
   for(j in 6:10){
     xb[i,j] <- b0[j] + b1[j] * x.pop[i]
     y[i,j] ~ dnorm(xb[i,j], tau2)
     y.star[i,j] ~ dnorm(xb[i,j], tau2)
    }

   # GDP Per Capita items
   for(j in 11:16){
     xb[i,j] <- b0[j] + b1[j] * x.gdppc[i]
     y[i,j] ~ dnorm(xb[i,j], tau3)
     y.star[i,j] ~ dnorm(xb[i,j], tau3)
    }

   # latent traits
   x.gdp[i] <- mu1[country[i], year[i]]
   x.pop[i] <- mu2[country[i], year[i]]
   x.gdppc[i] <- log(exp(x.gdp[i]) / exp(x.pop[i]))
 }

 # priors for the latent traits
 sigma[1] ~ dunif(0,1)
 kappa[1] <- pow(sigma[1], -1)  # invert to precision

 sigma[2] ~ dunif(0,1)
 kappa[2] <- pow(sigma[2], -1)  # invert to precision

 for(c in 1:n.country){
  mu1[c, 1] ~ dnorm(0, 1)
  mu2[c, 1] ~ dnorm(0, 1)
   for(t in 2:n.year){ #n.year is number of years
    mu1[c, t] ~ dnorm(mu1[c, t-1], kappa[1])
    mu2[c, t] ~ dnorm(mu2[c, t-1], kappa[2])
   }
  }

 # priors for the model estimates
 for(j in 1:k){
  b0[j] ~ dnorm(B0[j],4)
  b1[j] <- 1
  }

 tau1 ~ dgamma(0.001,0.001)
 tau2 ~ dgamma(0.001,0.001)
 tau3 ~ dgamma(0.001,0.001)
}

}\end{Verbatim}
}

\end{document}